\documentclass[acmsmall]{acmart}

\newcommand{\system}{\text{FENCE}}
\newcommand\mal{\ensuremath{\mathsf{Malicious}}}
\newcommand\ben{\ensuremath{\mathsf{Benign}}}
\newcommand{\ffnn}{\text{FFNN}}
\newcommand{\myparagraph}[1]{\smallskip \noindent \textbf{#1.}}
\newcommand\mymax{\ensuremath{\mathsf{Max}}}
\newcommand\mymin{\ensuremath{\mathsf{Min}}}

\newcommand\myargmax{\ensuremath{\mathsf{argmax}}}
\newcommand\myavg{\ensuremath{\mathsf{Avg}}}
\newcommand\mytotal{\ensuremath{\mathsf{Total}}}

\newcommand\FeasibleSet{\ensuremath{\mathsf{Feasible\_Set}}}
\newcommand{\vct}[1]{\ensuremath{\boldsymbol{#1}}}
\newcommand\updatefamily{\ensuremath{\mathsf{UPDATE\_FAMILY}}}
\newcommand\updatedep{\ensuremath{\mathsf{UPDATE\_DEP}}}
\newcommand\initfam{\ensuremath{\mathsf{INIT\_FAMILY}}}
\newcommand\updatestat{\ensuremath{\mathsf{Update\_Stat}}}
\newcommand\updateratio{\ensuremath{\mathsf{Update\_Ratio}}}
\newcommand\updaterange{\ensuremath{\mathsf{Update\_Range}}}
\newcommand\updatenonlin{\ensuremath{\mathsf{Update\_NonLin}}}
\newcommand\updatelin{\ensuremath{\mathsf{Update\_Lin}}}
\newcommand\updatecomb{\ensuremath{\mathsf{Update\_Comb}}}

\newcommand\xsumbytes{\ensuremath{x^{\mathsf{tot}}_{\mathsf{bytes}}}}
\newcommand\xminbytes{\ensuremath{x^{\mathsf{min}}_{\mathsf{bytes}}}}
\newcommand\xmaxbytes{\ensuremath{x^{\mathsf{max}}_{\mathsf{bytes}}}}

\newcommand\xsumdur{\ensuremath{x^{\mathsf{tot}}_{\mathsf{dur}}}}
\newcommand\xmindur{\ensuremath{x^{\mathsf{min}}_{\mathsf{dur}}}}
\newcommand\xmaxdur{\ensuremath{x^{\mathsf{max}}_{\mathsf{dur}}}}

\newcommand\xsumpackets{\ensuremath{x^{\mathsf{tot}}_{\mathsf{packets}}}}
\newcommand\xminpackets{\ensuremath{x^{\mathsf{min}}_{\mathsf{packets}}}}
\newcommand\xmaxpackets{\ensuremath{x^{\mathsf{max}}_{\mathsf{packets}}}}

\DeclareMathOperator{\argmin}{arg\,min}

\usepackage{algorithm}
\usepackage{algorithmic}
\usepackage{subcaption}
\usepackage{stfloats}

\begin{document}

\acmJournal{TOPS}

\title{FENCE: Feasible Evasion Attacks on Neural Networks in Constrained Environments}

\author{Alesia Chernikova}
\affiliation{%
  \institution{Northeastern University}
  \country{Boston, MA, USA}}
\email{chernikova.a@northeastern.edu}

\author{Alina Oprea}
\affiliation{%
  \institution{Northeastern University}
  \country{Boston, MA, USA}
  }
\email{a.oprea@northeastern.edu}

\begin{abstract}
As advances in Deep Neural Networks (DNNs) demonstrate unprecedented levels of performance in many critical applications, their vulnerability to attacks is still an open question. 
We consider  evasion attacks at testing time against Deep Learning in constrained environments, in which dependencies between features need to be satisfied. These situations may arise naturally in tabular data or may be the result of feature engineering in specific application domains, such as threat detection in cyber security. We propose a general iterative gradient-based framework called \system\ for crafting evasion attacks that take into consideration the specifics of constrained domains and application requirements. We apply it against Feed-Forward Neural Networks trained for two cyber security applications: network traffic botnet classification and malicious domain classification, to generate feasible adversarial examples. We extensively evaluate the success rate and performance of our attacks, compare their  improvement over several baselines, and analyze factors that impact the attack success rate, including the optimization objective and the data imbalance. We show that with minimal effort (e.g., generating 12 additional network connections), an attacker can change the model's prediction from the Malicious class to  Benign and evade the classifier. We show that models trained on datasets with higher imbalance are more vulnerable to our \system\ attacks. Finally, we demonstrate the potential of performing adversarial training in constrained domains to increase the model resilience against these evasion attacks.
\end{abstract}

\maketitle

\section{Introduction}
\label{sec:intro}

Deep learning has reached high performance in machine learning (ML) tasks in a variety of application domains, including  image classification, speech recognition, and natural language processing (NLP). Many  applications already benefit from the deployment of ML for automating decisions and we expect a proliferation of ML in even more critical settings in the near future.  Still, research in adversarial machine learning showed that deep neural networks (DNNs) are not robust in face of adversarial attacks. The first adversarial attack against DNNs was an evasion attack, in which  an adversary creates  adversarial examples that minimally perturb testing samples and change the classifier's prediction~\cite{Szegedy14}.  Since the discovery of adversarial examples in computer vision, a lot of work on evasion attacks against ML classifiers at deployment time has been performed.  Most of these attacks have been demonstrated in continuous domains (i.e., image classification), in which features or image pixels can be modified arbitrarily to create the perturbations~\cite{Szegedy14,Biggio13,Goodfellow14,kurakin2016adversarial,Papernot17,Carlini17,madry2017towards,athalye2018obfuscated}.

ML has a lot of potential in other application domains, including cyber security, finance, and healthcare, in which the raw data is not directly suitable for learning and engineered features are  defined by domain experts to train DNN models. Additionally, in certain application domains such as network traffic classification used in cyber security, the raw data itself might exhibit  domain-specific constraints in the original input space. Therefore, techniques used for mounting evasion attacks in continuous domains will not respect the feature-space dependencies  in these applications  and new adversarial attacks need to be designed for such constrained domains.

In this paper we introduce a novel, general framework  for mounting evasion attacks against deep learning models in constrained application domains. Our framework is named \system\ ({\bf F}easible {\bf E}vasion Attacks on {\bf N}eural Networks in {\bf C}onstrained {\bf E}nvironments). \system\ generates feasible adversarial examples in constrained domains that rely either on feature engineering or naturally have domain-specific dependencies in the input space. \system\ supports a range of linear and non-linear dependencies in feature space and can be applied to any higher-level classification task whose data respects these constraints.  At the core of \system\ is an iterative optimization method that determines the feature of the maximum gradient of the attacker's objective at each iteration, identifies the family of features dependent on that feature, and modifies consistently all those features, while preserving an upper bound on the maximum distance from the original sample. At any time during the iterative procedure, the input data point is modified within the feasibility region, resulting in feasible adversarial examples. 
Existing evasion attacks in  constrained environments, such as PDF malware detection~\cite{Srndic14,Tong19}, malware classification~\cite{grosse2016adversarial,suciu2018exploring}, and network traffic classification ~\cite{alhajjar2020adversarial,granados2020realistic,abusnaina2019examining,han2020practical,sadeghzadeh2021adversarial} do not support the entire range of complex mathematical and domain-specific dependencies as our \system\ framework. Moreover, some of these attacks result in worse performance~\cite{alhajjar2020adversarial} or operate only in a specific domain~\cite{sadeghzadeh2021adversarial} or feature space~\cite{alhajjar2020adversarial,granados2020realistic}.

We demonstrate that \system\ can successfully evade the DNNs trained for two cyber security  applications: a malicious network traffic classifier using the CTU-13 botnet dataset~\cite{Garca2014AnEC}, and a malicious domain classifier using the MADE system~\cite{oprea2018made}. In both settings, \system\ generates feasible adversarial examples with small modification of the original testing sample. For instance, by adding 12 network connections and preserving the original malicious behavior, an attacker can change the classification prediction of a testing sample from \mal\ to \ben\ in the network traffic classifier. We perform detailed evaluation to demonstrate that our attacks perform better than several baselines and existing attacks. We show that the state-of-the-art Carlini-Wagner attack~\cite{Carlini17} designed for continuous  domains does not respect the feature-space dependencies of our security applications. We compare two optimization objectives in our \system\ framework, the Projected Gradient Descent (PGD)~\cite{madry2017towards} and the Penalty method~\cite{Carlini17}, and show the advantages of the PGD optimization. We also study the impact of data imbalance on the classifier robustness and show that models trained on datasets with higher imbalance, as is common in security applications, are more vulnerable.

We also consider attack models with minimum knowledge about the ML system, in which the attacker does not have information about the exact model architecture and hyperparameters. We test several approaches for performing the attacks through transferability from a surrogate model to the original one, using the \system\ framework. We observe that the evasion attacks generated for different DNN architectures transfer to the target DNN model with slightly lower success than attacking directly the target model with \system. Finally, we test the resilience of adversarial training using our attacks as a defensive mechanism for DNNs trained in constrained environments.

To summarize, our contributions are:
\begin{enumerate}
    \item We introduce a general evasion attack framework \system\ for constrained application domains that supports a range of mathematical dependencies in feature space and two optimization approaches.
    \item We apply \system\ to two cyber security applications using different datasets and feature representations: a malicious network connection classifier, and a malicious domain detector, to generate feasible adversarial examples in these domains.
    \item We extensively evaluate \system\ for these applications, compare our attacks with several baselines, and quantify the adversarial success at different perturbations. We also study the impact of data imbalance on the classifiers' robustness. 
    \item We  evaluate the transferability of the proposed evasion attacks between different ML models and architectures, and show that adversarially-trained models provide higher robustness.
\end{enumerate}

\section{Background}
\label{sec:background}
\subsection{Deep Neural Networks for Classification}
A feed-forward neural network (\ffnn) for binary classification is a function $y = F(x)$ from input  $x \in R^d$ (of dimension $d$) to output $y \in \{0,1\}$. The parameter vector of the function  is learned during the training phase  using back propagation over the network layers. Each layer includes a matrix multiplication and non-linear activation (e.g., ReLU). The last layer's activation is sigmoid $\sigma$ for binary classification: $y=F(x) = \sigma(Z(x))$, where $Z(x)$ are the \emph{logits}, i.e., the output of the  penultimate layer. We denote by $C(x)$ the predicted class for $x$. For multi-class classification, the last layer uses a softmax activation function. There are other DNN architectures for classification, such as convolutional neural networks, but in this paper we  only consider \ffnn\ architectures.

\begin{figure*}[t]
\begin{centering}
\includegraphics[width=\linewidth]{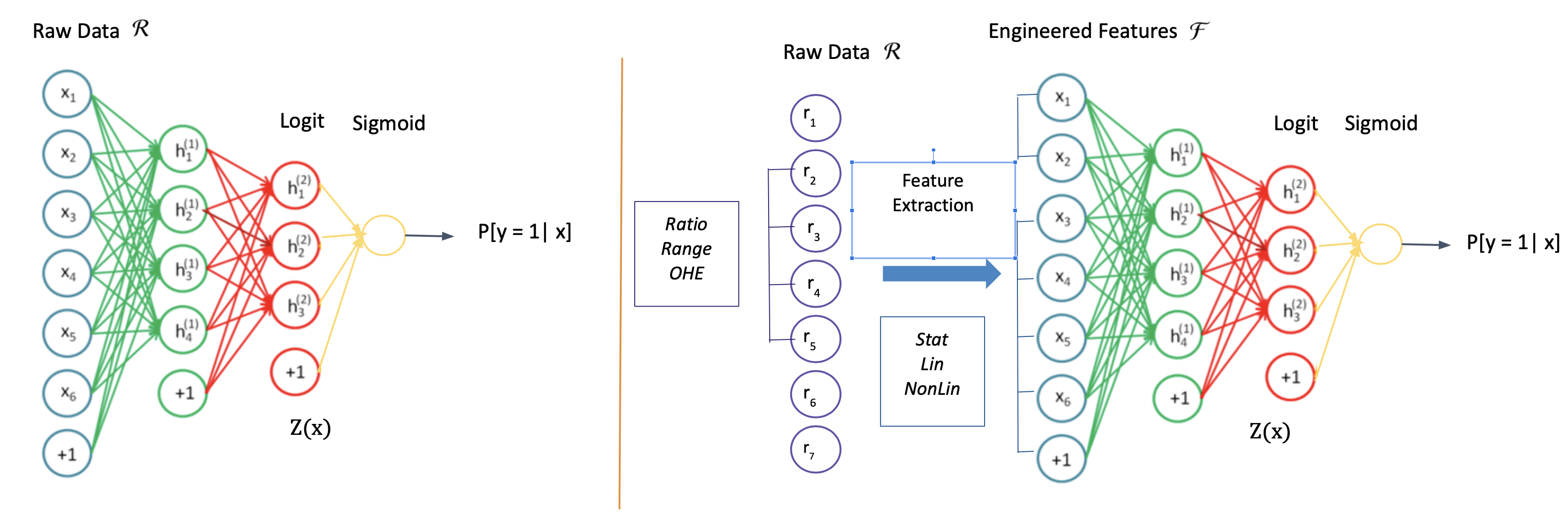}
\caption{Neural network training for images (left) and for constrained domains with  feature space dependencies (right). In the vision domain the raw data space $\mathcal{R}$ is the same as the feature representation $\mathcal{F}$, while in constrained domains they might be different. Additional dependencies in feature space might arise from natural dependencies in the raw data and the feature engineering process.}
\label{fig:learning_diagram}
\end{centering}
\end{figure*}

\subsection{Threat Model}
Adversarial attacks against ML algorithms can be developed in the training or testing phase. In this work, we consider testing-time attacks, called \emph{evasion attacks}. There exist several evasion attacks against DNNs in continuous domains: the projected gradient descent (PGD) attack~\cite{madry2017towards} and the penalty-based  attack of Carlini and Wagner~\cite{Carlini17}.

\myparagraph{Projected gradient attacks} This is a class of attacks based on gradient descent for objective minimization, that project the adversarial points to the feasible domain at each iteration. For instance, Biggio et al.~\cite{Biggio13}  use an objective that maximizes the confidence of adversarial examples, within a ball of fixed radius in $L_1$ norm. Madry et al.~\cite{madry2017towards} use the loss function directly as the optimization objective  and use the $L_2$ and $L_{\infty}$ distances for projection. 

\myparagraph{C\&W attack} Carlini and Wagner~\cite{Carlini17} solve the following optimization problem to create  adversarial examples against CNNs used for multi-class prediction:
\begin{center}
$\delta = \argmin ||\delta||_2+ c \cdot h(x + \delta)$\\
$h(x+\delta) = \max(0, \max(Z_k(x+\delta):k \neq t) - Z_t(x + \delta))$,\\
where $Z()$ are the logits of the \ffnn.
\end{center}

This is called the penalty method, and the optimization objective has two terms: the norm of the perturbation $\delta$, and a function $h(x + \delta)$ that is minimized when the adversarial example $x+\delta$ is classified as the target class $t$. The attack works for $L_0$, $L_2$, and $L_{\infty}$ norms.

Under the assumption that the DNN model is trained correctly, the attacker’s goal is to create adversarial examples at testing time. In security settings, typically the attacker starts with \mal\ points that he aims to minimally modify into adversarial examples classified as \ben. 

 Initially, we consider a  white-box attack model, in which the attacker has full knowledge of the ML system. White-box attacks have been considered extensively in previous work, e.g.,~\cite{Goodfellow14, Biggio13, Carlini17, madry2017towards} to evaluate the robustness of existing ML classification algorithms.  We also consider a more realistic attack model, in which the attacker has information about the feature representation of the underlying classifier, but no exact details on the ML algorithm and training data. 

We address application domains with various constraints in feature space. These could manifest directly in the raw data features or could be an artifact of the feature engineering process. The attacker has the ability to insert records in the raw data, for instance by inserting network connections in the threat detection applications. We ensure that the data points modified or added by the attacker are feasible in the constrained domain.
\section{Methodology}
\label{sec:method}

In this section, we start by describing the classification setting in constrained domains with dependencies in  feature space and the challenges of evasion attacks in this setting. Then we devote the majority of the section to present our new attack framework \system\ which takes into consideration the relationships between features that occur naturally in the problem space or are the result of feature engineering.

\subsection{Machine Learning Classification in Constrained Domains}

Let the raw data input space be denoted as $\mathcal{R}$. This is the original space in which raw data is collected for an application. In healthcare, $\mathcal{R}$ could be the space of all data collected for a particular patient. In network security, $\mathcal{R}$ could be the raw network traffic (for example, pcap files or Zeek network logs) collected in a monitored network in order to detect cyber attacks.

Consider a fixed  raw data set $R = \{r_1,\dots, r_M\} \in \mathcal{R}$. The raw data is typically processed into a feature representation, denoted by $\mathcal{F}$, over which the machine learning model is trained.  In standard computer vision tasks such as image classification, the raw data (image pixels) is used directly as input for neural networks. Thus, the training examples $x_i$ are the same as the raw data: $x_i = r_i, i \in [1,M]$. In this case the feature space $\mathcal{F}$ is the same as the input space $\mathcal{R}$.

In contrast, in other domains, such as threat detection or health care, the feature representation is not always exactly the raw data. See Figure~\ref{fig:learning_diagram} for a visual representation of this process. In most application domains,  there might exist dependencies and constraints in the feature space introduced either by the application itself or by the feature engineering process:
\begin{itemize}
    \item Dependencies among different features  could manifest naturally in the considered application. For instance, the results of two blood tests are correlated and they result in correlated feature values for a patient data. In network security, the packet size and number of packets are correlated with the total number of bytes sent in a TCP connection. We denote by $\FeasibleSet(R)$ the set of all feasible points in the raw data space. $\FeasibleSet(R)$ is a subset of the raw data that encompasses the feasible values for the particular application. For instance, a network TCP packet size is upper bounded  by 1500 bytes and the ratio between the number of bytes and the number of packets in a TCP connection needs to be lower than the maximum packet size.  The feasible set $\FeasibleSet(R)$ will only include network connections in which the upper bound on TCP packets is enforced and the ratio constraint between the number of bytes and number of packets  is satisfied.
    
    \item Constraints in feature representations might also result from the feature engineering process performed in many settings. In this case, features in $\mathcal{F}$ are obtained by the application of an operator  $\mathsf{Op}_j$ on the raw data  $R \in \mathcal{R}$: $x_{ij} = \mathsf{Op}_j (R)$. Examples of operators are statistical functions such as \mymax, \mymin, \myavg, and \mytotal, as well as linear combinations of raw data values, and other mathematical functions such as product or ratio of two values. The set of all supported operators applied to the raw data is denoted by $\mathcal{O}$. This process creates $N$ training examples $x_1,\dots,x_N$ in the feature space $\mathcal{F}$, each being $d$-dimensional, with $d$ the size of the feature space. The feature engineering process creates additional dependencies in feature space. For instance, if we consider the \mymax, \mymin, and \myavg\ number of connections for a particular port in a given time window, the average value needs to be between the minimum and the maximum values. 
 
\end{itemize}

A data point $z = (z_1,\dots,z_d)$ in feature space $\mathcal{F}$ is \emph{feasible} if there exists some raw data $R \in \mathcal{R}$ such as for all $i$, there exists an operator $\mathsf{Op}_j \in \mathcal{O}$ with $z_i = \mathsf{Op}_j (R)$. The set of all feasible points in feature space for raw data $R$ and operators $\mathcal{O}$ is called $\FeasibleSet(R,\mathcal{O})$. This space includes the set of feasible points $\FeasibleSet(R)$ (obtained for $\mathcal{O} = \emptyset$). Examples of feasible and infeasible points in feature space are illustrated in Table~\ref{tab:feasible}. The constraints in this example are that the sum  of  feature values  must sum up to one. This may arise in situations when the subset of features represents ratio values, for example, the ratio of connections that have a particular result code.

\begin{table}[h]
\centering
\begin{tabular}{|c||c||c|}
\hline
Feature & Feasible point & Infeasible point\\
\hline
$F_1$ & 0.2 &  {\color{red} 0.5}\\
$F_2$ & 0.13 & {\color{red} 0.13}  \\
$F_3$ & 0.33 & {\color{red} 0.33} \\
$F_4$ & 0.34 & {\color{red} 0.4} \\
\hline
\end{tabular}
\caption{Example feature values for four ratio features, whose sum needs to be 1 in the feasibility region. We show an example of a feasible point and an infeasible one in feature space $\mathcal{F}$.}
\label{tab:feasible}
\end{table}

\begin{figure*}[ht]
\begin{centering}
\includegraphics[width=\linewidth]{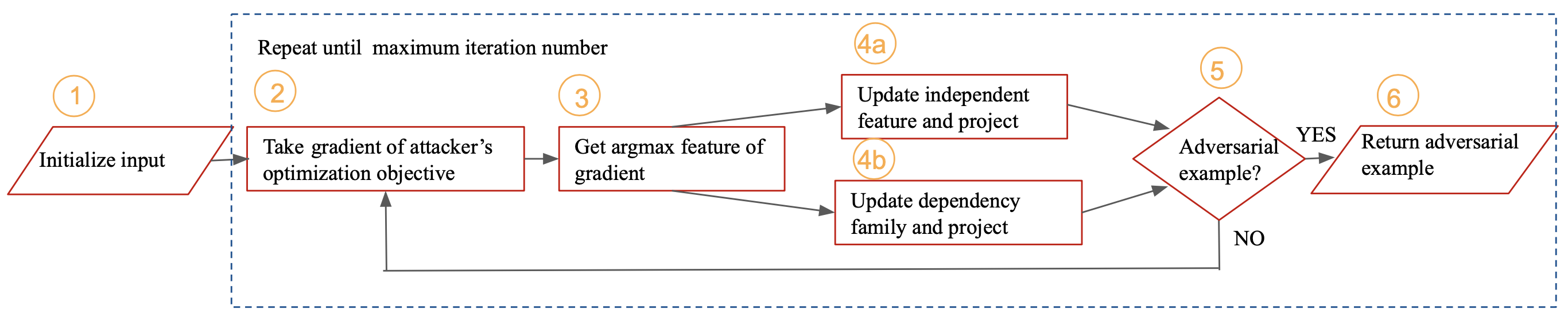}
\caption{Flow of the \system\ Evasion Attack Framework.}
\label{fig:ea_diagram}
\end{centering}
\end{figure*}

\subsection{Challenges}
Existing evasion attacks are mostly designed for continuous domains, such as image classification, where adversarial examples have pixel values in a fixed range (e.g., [0,1]) and can be modified independently~\cite{Carlini17,madry2017towards,athalye2018obfuscated}. However, many applications in cyber security use tabular data, resulting in feature dependencies and physical-world constraints that need to be respected.

Several previous works address evasion attacks in domains with tabular data.  The evasion attack for malware detection by Grosse et al.~\cite{Grosse17}, which directly leverages JSMA~\cite{Papernot17}, modifies binary features corresponding to system calls. 
Kolosnjaji et al.~\cite{kolosnjaji2018adversarial} use the attack of Biggio et al.~\cite{Biggio13} to append selected bytes at the end of the malware file. Suciu et al.~\cite{suciu2018exploring} also append bytes in selected regions of malicious files. Kulynych et al. ~\cite{kulynych2018evading} introduce a graphical framework in which an adversary constructs all feasible transformations of an input, and then uses graph search to determine the path of the minimum cost to generate an adversarial example. 

 Neither of these approaches is applicable to our general setting because the attacks do not satisfy the required dependencies in the resulting adversarial vector. Crafting adversarial examples that are feasible, and respecting all the application constraints and dependencies poses a significant challenge. 
 Once application constraints are specified, the resulting optimization problem for creating adversarial examples includes a number of non-linear constraints and cannot be solved directly using out-of-the-box optimization methods.

In order to measure the feasibility of adversarial examples, we run the existing Carlini and Wagner (C\&W) attack~\cite{Carlini17} on a malicious domain classification. The details of the attack adaptation  for this application are given in Section~\ref{sec:madeapp}. We considered the balanced case, in which the number of \mal\ and \ben\ examples is equal in training. While the attack reaches 98\% success at a distance of 20, the resulting adversarial examples are outside the feasibility region. An example is included in Table~\ref{tab:cw}, and the description of the features is given in Table~\ref{tab:features_id}. In this case,  the average number of connections is not equal to the total number of connections divided by the number of IPs contacting the domain. Additionally, the average ratio of received bytes over sent bytes should be equal to  the maximum and minimum values (as the number of IPs contacting the domain is 1).

\begin{table}[ht]
\centering
\begin{tabular}{|c||c||c||c||c|}
\hline
Feature & Description & Input & Adversarial Example& Correct Value\\
\hline
Num\_IP & Number of IPs & 1 & 1 & 1\\
Num\_Conn & Number of connections & 15 &233.56 & 233.56\\
Avg\_Conns  & Avg. connections by IP & 15 & \color{red}59.94 &\color{green} 233.56 \\
Avg\_Ratio\_Bytes & Avg. ratio bytes &8.27&204.01 &204.01  \\
Max\_Ratio\_Bytes & Max. ratio bytes & 8.27 & \color{red}240.02 &\color{green}204.01  \\
Min\_Ratio\_Bytes & Min. ratio of bytes & 8.27&\color{red}119.12& \color{green}204.01 \\
\hline
\end{tabular}
\caption{Infeasible C\&W adversarial example for the malicious domain classifier. The inconsistent feature values are in red, while the correct values are in green. We consider all the IPs that connect to the domain and all the connections made to the domain. The ratio of bytes features are computed as ratio of received bytes over sent bytes per IP. When there is a single IP connecting to the domain, the number of connections should be equal to the average number of connections by IP, and the three ratio bytes features should  all be equal.}
\label{tab:cw}
\end{table}

\subsection{The \system\ framework} 

To address these issues, we introduce the \system\ framework for evasion attacks that preserves a range of feature dependencies in constrained domains. \system\ guarantees by design that the produced adversarial examples are within the feasible region of the application input space. 

The starting point for the attack framework are gradient-based optimization algorithms, including projected~\cite{Biggio13,madry2017towards} and penalty-based~\cite{Carlini17} optimization methods. Of course, we cannot apply these attacks directly since they will not preserve the feature dependencies. To overcome this, we use the values of the objective gradient at each iteration to select features of maximum gradient values. We create feature-update algorithms for each family of dependencies that use a combination of gradient-based method and mathematical constraints to always maintain a feasible point that satisfies the constraints. We also use various projection operators to project the updated adversarial examples to feasible regions of the feature space. 

\begin{algorithm}[th]
    \caption{\system\ Framework for Evasion Attack with Constraints}
	\label{alg:evasion_fw}
	\begin{algorithmic}[1]
		\REQUIRE$\vct x, y$: the input sample and its label; 
		$t$: target label;
		$C$: prediction function; 
		$G$:  optimization objective; 
		$d_{max}$: maximum allowed perturbation; 
		$F_S$: subset of features that can be modified 
		$F_D$: features in $F_S$ that have dependencies; 
		$M$: maximum number of iterations; 
		$\alpha$: learning rate.
		\ENSURE $\vct x^{*}$: adversarial example or $\bot$ if not successful.
		\STATE Initialize $m \leftarrow 0; x^0 \leftarrow \vct x$
		\STATE // Iterate until successful or stopping condition
		\WHILE {$C(x^m) != t$ and $m < M$ }
			\STATE $\nabla \leftarrow  [\nabla G_{x_i}(x ^m)]_i$ // Gradient vector
			\STATE $\nabla_S \leftarrow   \nabla_{F_S}$ // Gradients of features in $F_S$
			\STATE $ i_{max} \leftarrow \myargmax \nabla_S $ // Feature of max gradient 
			\STATE // Check if feature has dependencies
			\IF{$ i_{max} \in F_D$} 
				\STATE // Update dependent features
				\STATE $x^{m+1} \leftarrow \updatefamily(m,x^m, \nabla, i_{max})$
			\ELSE 
				\STATE Gradient update and projection 
				\STATE $x_{i_{max}}^{m+1} \leftarrow x^m_{i_{max}} - \alpha \nabla_{i_{max}}$
				\STATE $x^{m+1} \leftarrow \Pi_2(x^{m+1})$
			\ENDIF
			\STATE $F_S \leftarrow F_S \setminus \{i_{max}\}$
			\STATE $m \leftarrow m+1$
			\IF{$C(x^m)=t$} 
			    \STATE $x^* \leftarrow$ PROJECT\_TO\_RAW($x^m$)
			    \RETURN $x^*$
			\ENDIF
		\ENDWHILE
		\RETURN $\bot$
	\end{algorithmic}
\end{algorithm}
\begin{algorithm}[th]
\caption{\updatefamily\ ($m,x^m, \nabla, i_{max}$)}
    \label{alg:evasion_fw_upd}
	\begin{algorithmic}[1]
			\STATE // Extract all dependent features on $i_{max}$	
			\STATE $F_{i_{max}} \leftarrow \mathsf{Family\_Dep}(i_{max})$
			\STATE // Family representative feature 
			\STATE $j \leftarrow \mathsf{Family\_Rep}(F_{i_{max}})$ 		
			\STATE $\delta \leftarrow \nabla_j $ // Gradient of representative feature
			\STATE // Initialization function  
			\STATE $s \leftarrow \initfam(x^m, \nabla, j)$ 
			\STATE // Binary search for perturbation 
			\WHILE{$\delta \neq 0$}
				\STATE $x^m_j \leftarrow x^m_j - \alpha \delta$ // Gradient update
				\STATE $x^m \leftarrow \updatedep(s,x^m,\nabla,F_{i_{max}})$
				\IF{$d(x^m, x^0) > d_{max}$}
					\STATE // Reduce perturbation
					\STATE $\delta \leftarrow \delta / 2$
				\ELSE
					\RETURN $x^m$
				\ENDIF
			\ENDWHILE
	\end{algorithmic}
\end{algorithm}
Algorithms~\ref{alg:evasion_fw}, \ref{alg:evasion_fw_upd} and Figure~\ref{fig:ea_diagram} describe the general \system\ framework. We consider binary classifiers designed using \ffnn\ architectures. However, the framework can be extended to multi-class scenarios by modifying the optimization objective. For measuring the amount of perturbation added by the original example, we use the $L_2$ norm.  

The input to the \system\ framework consists of: an input sample $\vct x$ with label $y$ (typically \mal\ in security applications); a target label $t$ (typically \ben); the model prediction function $C$; the optimization objective $G$; maximum allowed perturbation $d_{max}$; the subset of features $F_S$ that can be modified; the features that have dependencies $F_D \subset F_S$; the maximum number of iterations $M$ and a learning rate $\alpha$ for gradient descent. The set of dependent features is split into families of features. A family is defined as a subset of $F_D$ such that features within the family need to be updated simultaneously, whereas features outside the family can be updated independently. In our malicious network traffic classification application, a family of features is defined for each port by including all features extracted for that particular port. All these features are dependent and they are modified jointly during the adversarial optimization procedure. 

The algorithm proceeds iteratively. The goal is to update the data point in the direction of the gradient (to minimize the optimization objective) while preserving the domain-specific and mathematical dependencies between features. 
In each iteration, the gradients of all modifiable features are computed, and the feature of the maximum gradient is selected. The update of the data point $x$ in the direction of the gradient is performed as follows:

 \begin{enumerate}
     \item 
 If the feature of maximum gradient belongs to a family with other dependent features, function \updatefamily\ is called (Algorithm~\ref{alg:evasion_fw}, line 10). Inside the function, the representative feature for the family is computed (this needs to be defined for each application). In the malicious network traffic classification example, the representative feature for a port's family is the number of sent packets on that port.  The representative feature is updated first, according to its gradient value, followed by updates to other dependent features using the function \updatedep\ (Algorithm~\ref{alg:evasion_fw_upd}, line 11). We need to define the function \updatedep\ for each application, but we use a set of building blocks that support common operations in feature space and are reusable across applications. We refer the reader to Section~\ref{sec:dep} for a set of dependencies supported by our framework. Once all features in the family have been updated, there is a possibility that the updated data point exceeds the allowed distance threshold from the original point. If that is the case, the algorithm backtracks and performs a binary search for the amount of perturbation added to the representative feature (until it finds a value for which the modified data point is inside the allowed region). 

\item If the feature of maximum gradient does not belong to any feature family, then it can be updated independently from other features. The feature is updated using the standard gradient update rule (Algorithm~\ref{alg:evasion_fw}, line 13). This is followed by a projection $\Pi_2$ within the feasible $L_2$ ball. 
\end{enumerate}

Finally, if the attacker is successful at identifying an adversarial example in feature space, it is projected back to the raw input  space representation (using function PROJECT\_TO\_RAW). With the \system\ framework, we guarantee that modifications in feature space always result in feasible regions of the feature space, meaning that they can be projected back to the raw input space. Of course, if the raw data representation is used directly as features for the ML classifier, the projection is not necessary.

\system\ currently supports two optimization objectives: 

\begin{itemize}
    \item 
\emph{Objective for Projected attack.} We set the objective $G(x)=Z_1(x)$, where $Z_1$ is the logit for the \mal\ class, and $Z_0 = 1-Z_1$ for the \ben\ class:

\begin{center}
$\delta = \argmin Z_1(x+\delta)$,\\
s.t. $||\delta||_2 \leq d_{max}$, \\
$x+\delta \in \FeasibleSet(R,\mathcal{O})$
\end{center}

\item
\emph{Objective for Penalty attack.} The penalty objective for binary classification is equivalent to:

\begin{center}
$\delta = \argmin ||\delta||_2+ c \cdot \max(0, Z_1(x + \delta))$,\\
$x+\delta \in \FeasibleSet(R,\mathcal{O})$
\end{center}

\end{itemize}

Our general \system\ evasion framework can be used for different classifiers, with multiple features representations and constraints. The components that need to be defined for each application are: (1) the optimization objective $G$ for computing adversarial examples; (2) the families of dependent features and family representatives; (3) the \updatedep\ function that performs feature updates per family; (4) the projection operation  PROJECT\_TO\_RAW that transforms adversarial examples from feature space to the raw data input. 

\subsection{ Dependencies in Feature Space}
\label{sec:dep}
In this section, we describe the dependencies in the feature space that \system\ supports. For each of these, there is a corresponding \updatedep\ algorithm used in the \system\ optimization framework. Once the representative feature in a family is updated according to the gradient value in Algorithm~\ref{alg:evasion_fw}, the dependent features are updated with the \updatedep\ algorithm.

\myparagraph{Domain-Specific Dependencies} The supported domain-specific dependencies are illustrated in  Table~\ref{tab:dsr}. These dependencies might occur naturally in the raw data space. The \emph{Range} dependency ensures that feature values are in a particular numerical range, while the \emph{Ratio} dependency  ensures that the ratio of two features is in a particular interval.  The one-hot encoded feature dependency is a  structural dependency of the input vector representation, encountered when categorical data is represented by creating a binary feature for each value. Algorithms ~\ref{alg:dsd} and ~\ref{alg:rangeud} describe how to preserve the \emph{Range} and \emph{Ratio} dependencies, respectively. 

Algorithm~\ref{alg:dsd} illustrates the procedure for updating dependent features to satisfy the \emph{Ratio} relationship. If the dependency between two features $x$ and $y$ is such that $x/y \in [a, b]$, then feature $x$ is modified according to the gradient value, but the final range is restricted to the interval $[a \cdot y,b \cdot y]$.

Algorithm~\ref{alg:rangeud} gives the update function for \emph{Range}. It ensures that input $x$ is projected to interval $[a,b]$. It returns the projected value of $x$, as well as the absolute value of the difference between $x$ and its projection.

\begin{table}[ht]
\centering
\begin{tabular}{|c||c|}
\hline
Type of dependency & Formula\\
\hline
$Range$ &$x_i: x_i \in [a, b]$ \\
\hline
$Ratio$  &$x_i, x_j: x_i/x_j \in [a,b]$\\
\hline
One-hot encoding (\textbf{$OHE$})&$\{x_i\}_1^N: x_i \in \{0, 1\}, \sum_{i = 1}^N{x_i} = 1$\\
\hline
\end{tabular}
\caption{Domain-specific feature dependencies.}
\label{tab:dsr}
\end{table}

\myparagraph{Mathematical Feature Dependencies}  Mathematical dependencies resulting from  feature engineering supported by \system\ are illustrated in  Table~\ref{tab:fsd}. These include statistical dependencies, linear and non-linear dependencies between multiple features, as well as combinations of these. To provide some insight, Algorithm~\ref{alg:efd} and Algorithm~\ref{alg:statudf} illustrate how to preserve $NonLin$ and $Stat$ dependencies.

Algorithm~\ref{alg:efd} shows the \emph{NonLin} update feature dependency procedure. Here, we need to ensure that the constraint $x_i - x_j / x_k = 0$ is satisfied for three features $x_i,x_j$, and $x_k$. Gradient update is performed first for $x_j$, after which the value of $x_i$ is modified to ensure the equality constraint, while feature $x_k$ is kept constant.

Algorithm~\ref{alg:statudf} gives the update method for satisfying the \emph{Stat} dependency. This is done for a family of features that includes the minimum $x_{min}$, the average $x_{avg}$, the maximum $x_{max}$, and the total number $x_{tot}$ from some events from the raw data. After the update of feature $x_{tot}$ (by increasing, for example, the total number of network connections), we need to adjust the average value  $x_{avg}$ and the corresponding minimum and maximum values. The input to \updatestat\ also includes a value $v$ that is the new value added to the raw data, which could impact the minimum or maximum values.

\begin{table}[ht]
\centering
\begin{tabular}{|c||c|}
\hline
Type of dependency & Formula\\
\hline
Statistical ($Stat$) & $ x_{min} \leq x_{avg}\leq x_{max} $ \\
\hline
Linear ($Lin$)& $\sum_i^M (w_i * x_i) = Ct$\\
\hline
Non-linear ($NonLin$) & $x_i - x_j/x_k = 0$\\
\hline
Combinations of $Lin$, & $x_{min} \leq (x_j/x_k)_{avg}\leq x_{max}$\\
 $Stat$, and $NonLin$ & $ \sum_i^M (w_i * x_i/x_k) = Ct$\\
\hline
\end{tabular}
\caption{Mathematical feature dependencies.}
\label{tab:fsd}
\end{table}


\begin{algorithm}[ht]
	\caption{\updateratio\ ($x,\nabla,F$)}
	\label{alg:dsd}
	\begin{algorithmic}[1]
		\STATE Parse $F$ as $a, b, x, y$ such that $x \in [a \cdot y,b \cdot y]$.
		\IF {$x - \alpha \cdot \nabla_x < a \cdot y$}
		    \STATE $x' \leftarrow a \cdot y$
		\ELSE \IF{ $x - \alpha \cdot \nabla_x > b \cdot y$}
		        \STATE $x' \leftarrow  b \cdot y$
		      \ENDIF
		\ENDIF
		\STATE $x' \leftarrow x -\alpha \cdot \nabla_x$
		\RETURN $x'$
	\end{algorithmic}
\end{algorithm}

\begin{algorithm}[ht]
	\caption{\updaterange\ ($x, a, b$)}
	\label{alg:rangeud}
	\begin{algorithmic}[1]
		\STATE $x' \leftarrow x$
		\IF {$x < a$}
		    \STATE $x' \leftarrow a$
		    \ENDIF
		\IF{ $x > b$}
		        \STATE $x' \leftarrow  b$
		      \ENDIF
		\RETURN $x'$, $|x' - x|$
	\end{algorithmic}
\end{algorithm}

\begin{algorithm}[ht]
	\caption{\updatenonlin\ ($x,\nabla,F$)}
	\label{alg:efd}
	\begin{algorithmic}[1]
	\STATE Parse $F$, the family of dependencies as: $x_i,x_j,x_k$. 
	\STATE $x'_j \leftarrow x_j - \alpha \nabla_j$
	\STATE $x'_i \leftarrow  x'_j/x_k$
	\STATE $x_j \leftarrow x'_j$, $x_i \leftarrow x'_i$
	\RETURN $x'$
	\end{algorithmic}
\end{algorithm}

\begin{algorithm}[ht]
	\caption{\updatestat\ ($x, v, F$)}
	\label{alg:statudf}
	\begin{algorithmic}[1]
	\STATE Parse $F$, the family of dependencies as $x_{min}, x_{max}, x_{avg}, x_{tot}, x_{num}$.
	\STATE $x'_{avg} \leftarrow  x_{tot} / x_{num}$ 
    \STATE $x'_{min} \leftarrow \mymin(x_{min}, v )$
	\STATE $x'_{max} \leftarrow \mymax(x_{max},  v)$
	\RETURN $x'$
	\end{algorithmic}
\end{algorithm}

\section{Concrete Applications of \system}
\label{sec:attacks}

In this section we describe the  application of \system\ to two classification problems for threat detection: malicious network traffic classification, and malicious domain classification. We highlight that \system\ can be applied to other domains with feature constraints such as healthcare and finance, but our focus in the paper is on cyber security applications.

\subsection{Malicious Network Traffic Classification}
\label{sec:nerisattack}

Network traffic includes important information about communication patterns between source and destination IP addresses. Classification methods have been applied to labeled network connections to determine malicious infections, such as those generated by botnets~\cite{EXPOSURE,Bartos16,BAYWATCH,oprea2018made}. Network data comes in a variety of formats, but the most common include net flows, Zeek logs, and packet captures. 

\myparagraph{Dataset} 
We leverage a public dataset of botnet traffic that was captured at the CTU University in the Czech Republic, called the CTU-13 dataset~\cite{Garca2014AnEC}. We consider the three scenarios for detecting the Neris botnet in this dataset. The dataset includes Zeek connection logs with communications between internal IP addresses (on the campus network) and external ones. The dataset has the advantage of providing ground truth, i.e., labels of \mal\ and \ben\ IP addresses. The goal of the classifier is to distinguish \mal\ and \ben\ IP addresses on the internal network. 

\begin{figure*}[t]
\begin{centering}
\includegraphics[width=\linewidth]{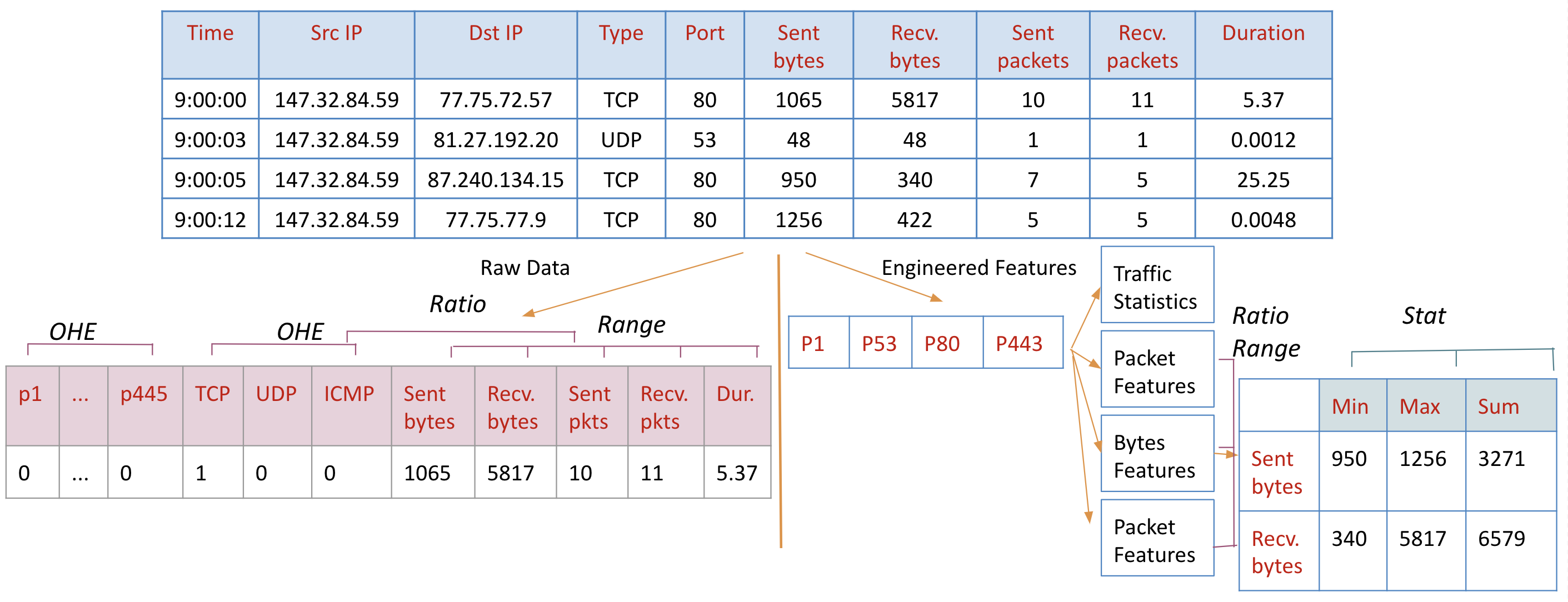}
\caption{Zeek logs (top), raw data representation (left), and feature families per port (right) for network traffic classifier.}
\label{fig:bro_example}
\end{centering}
\end{figure*}

The fields available in Zeek connection logs are given in Figure~\ref{fig:bro_example}. They include: the \emph{timestamp} of the connection start; the \emph{source IP address}; the \emph{source port}; the \emph{destination IP address}; the \emph{destination port}; the \emph{number of packets sent and received}; the \emph{number of bytes sent and received}; and the connection \emph{duration} (the time difference between when the last packet and first packets are sent).

In this application, we can use  either the raw connection representation or leverage domain knowledge to create aggregated features. We describe existing feature relationships and apply our \system\ framework against both representations. 

\myparagraph{Raw Data Representation}
This consists of the following fields: \emph{one-hot encoded port number}, \emph{one-hot encoded connection type}, \emph{duration}, \emph{original bytes}, \emph{received bytes}, \emph{original packets}, and \emph{received packets}. The feature vector is illustrated in Figure \ref{fig:bro_example} on the left.
The raw data representation  includes no mathematical dependencies, but has the following \textbf{domain-specific constraints}:

\vspace{1pt}
- The TCP and UDP packet sizes are capped at 1500 bytes. We create range intervals for these values, resulting in a $Ratio$ dependency between the number of packets and their sizes.

\vspace{1pt}
- The connection duration is the interval between the last and the first packet. If the connection is idle for some time interval (e.g., 30 seconds), it is closed by default by Zeek. The attacker can thus control the duration of the connection by sending packets at certain time intervals to avoid closing the connection. We generate a range of valid durations from the distribution of connection duration in the training dataset. This creates again a $Ratio$ dependency between the number of packets and their duration.

\vspace{1pt}
- Each continuous feature  has its related minimum and maximum values, which are obtained from the training data distribution, thus forming  $Range$ relationships.

\vspace{1pt}
- The port number and connection type  have one-hot encoded $OHE$ dependencies.

\myparagraph{Attack algorithm on raw data representation} The attacker's goal is to have a connection log classified as \ben\ instead of \mal. We assume that the attacker communicates with an external IP under its control (for instance, the command-and-control IP), and thus has full control of the malicious traffic in that connection. We assume that the attacker \emph{can only add traffic to network connections}, by increasing the number of bytes, packets, and connection duration, to preserve the malicious functionality. For simplicity, we set the number of received packets and bytes to 0, assuming that the  external IP does not respond to these connections. We assume that the attacker does not have access to the security monitor that collects the logs and cannot modify directly the log data. 

The attack algorithm follows the framework from Algorithm~\ref{alg:evasion_fw}. There is only one family of dependent features, including the packets and bytes sent, and connection duration.
The representative feature is the number of sent packets, which is updated with the gradient value, following a binary search for perturbation $\delta$, as specified in Algorithm \updatefamily. The dependent number of bytes sent and duration features are updated using the update dependency functions (\updateratio, Update\_Range and Update\_OHE), thus preserving the
$Ratio$, $Range$ and $OHE$ dependencies. 

\myparagraph{Engineered Features}
Another possibility is to use domain knowledge to create features that improve classification accuracy. A standard method for creating network traffic features  is aggregation by destination port to capture relevant traffic statistics per port (e.g.,~\cite{Garca2014AnEC}, \cite{ongun2019designing}). This is motivated by the fact that different network services and protocols run on different ports, and we expect ports to have different traffic patterns. We select a list of 17 ports for popular applications, including HTTP (80), SSH (22), and DNS (53). We also add a category called OTHER for connections on other ports. We aggregate the communication on a port based on a fixed time window (the length of which is a hyper-parameter set at one minute). For each port, we compute traffic statistics using the \mymax, \mymin, and \mytotal\ operators for outgoing and incoming connections. See the example in Figure~\ref{fig:bro_example} on the right, in which features extracted for each port define a family of dependent features. We obtain a total of 756 aggregated traffic features on these 17 ports. Table~\ref{tab:features_id_neris} includes the feature description.
The resulting feature vector includes both types of dependencies. The \textbf{domain-specific relationships} are the same as for the raw data representation except for the one-hot encoding  relationship. There are additional $Stat$ \textbf{mathematical dependencies} between features: the minimum and the maximum number of packets, bytes and duration per connection must be updated after a change in the total number of packets, bytes, or connections.

\begin{table*}
\centering
\begin{tabular}{|c||c||c|}
\hline
Category & Feature  & Description\\ 
\hline
Bytes & Total\_Sent\_Bytes & Total number of bytes sent \\
&Min\_Sent\_Bytes & Minimum number of bytes sent per connection\\
&Max\_Sent\_bytes & Maximum of bytes sent per connection\\
\hline
Packets& Total\_Sent\_Pkts & Total number of packets sent\\
&Min\_Sent\_Pkts & Minimum number of packets sent per connection\\
&Max\_Sent\_Pkts & Maximum of packets sent per connection\\
\hline
Duration& Total\_Duration&Total duration of all connections \\
&Min\_Duration&Minimum duration of a connection\\
&Max\_Duration&Maximum duration of a  connection\\
\hline
Connection type & Total\_TCP & Total number of TCP connections\\
& Total\_UDP & Total number of UDP connections\\
\hline
\end{tabular}
\caption{Features definition for malicious connection classification. These features are defined  for each port by aggregating over all connections on that port in a fixed time window.}
\label{tab:features_id_neris}
\end{table*}

\myparagraph{Attack algorithm on engineered features}  The goal of the attacker here is to change the prediction of a feature vector aggregated over time from \ben\ to \mal. Therefore, in this attack model, the attacker has the ability \emph{to insert network connections}  during the targeted time window to achieve his goal. Similar to the above scenario, the attacker controls a victim IP and can send traffic to external IPs under its control. The adversary has a lot of options in mounting the attack by selecting the protocol, port, and connection features. Here we have 17 families of dependent features, one for the features on each port.

The attack algorithm against the Neris botnet classification task called the Neris attack follows the framework from Algorithm~\ref{alg:evasion_fw}. First, the feature of the maximum gradient is determined and the corresponding port is identified. The family of  dependent features is all the features computed for that port. The attacker attempts to add a fixed number of connections on that port (which is a hyper-parameter of our system). This is done in the \initfam\ function. The attacker can add either TCP, UDP, or both types of connections, according to the gradient sign for these features and also respecting network-level constraints. The representative feature for a port's family is the number of packets that the attacker sends in a connection. This feature is updated by the gradient value, following a binary search for perturbation $\delta$, as specified in Algorithm \updatefamily\ .

In the \updatedep\ function an update to the aggregated port features is performed. First, we ensure that the feature corresponding to the number of packets sent satisfies the $Range$ and $Stat$ constraints. Then the difference in the total number of bytes sent by the attacker and duration is determined from the gradient, followed by the \updateratio\ function to keep the resulting values inside the feasible domain. The port family also includes features such as \mymin\ and \mymax\ sent bytes and connection duration. These features are updated by the \updatestat\ function. The detailed algorithm for \initfam\ and \updatedep\ functions of the Neris attack are illustrated in Algorithm~\ref{alg:neris_init} and Algorithm~\ref{alg:neris}.

\begin{algorithm}[th]
	\caption{Neris 	\initfam\ ($x^m, \nabla, j$)} 
	\label{alg:neris_init}
	\begin{algorithmic}[1]
	\REQUIRE $x^m$: data point in iteration $m$\\
	$\nabla$: gradient of objective with respect to $x$ \\
	$j$: representative feature of the the family\\
	$p$: port updated in iteration $m$\\
	$c$: number of connections to add\\
	    \STATE // Check if the connections are allowed on port $p$
		\STATE $ac \leftarrow$ ALLOWED\_CONNECTIONS$(p, x^m)$
		\IF{$\nabla_c <0$ and $ac == True$}
			\RETURN $c$
		\ENDIF
		\RETURN 0
	\end{algorithmic}
\end{algorithm}
\begin{algorithm}[th]
	\caption{Neris \updatedep\ ($s, x^m,\nabla, F_{i_{max}}$)} 
	\label{alg:neris}
	\begin{algorithmic}[1]	
	\REQUIRE $s$: number of added connections\\
	$x^m$: data point in iteration $m$\\
	$F_{i_{max}}$: all dependent features on the feature of maximum gradient $i_{max}$ \\
	\xsumpackets/\xminpackets/\xmaxpackets: total/min/max number of sent packets on $p$ per connection\\
	\xsumbytes/\xminbytes/\xmaxbytes: total/min/max number of sent bytes on $p$ per connection\\
	\xsumdur/\xmindur/\xmaxdur: total/min/max duration on $p$ per connection\\
	$\nabla$: gradient of objective with respect to $x$ \\
	$minP$: the minimum total number of sent packets from data distrinution\\
	$maxP$: the maximum total number of sent packets from data distribution\\
	$minB$: the minimum total number of sent bytes from data distrinution\\
	$maxB$: the maximum total number of sent bytes from data distribution\\
	$minD$: the minimum total connections duration from data distribution\\
	$maxD$: the maximum total connections duration from data distribution
	    \STATE $x^{\mathsf{tot}}_{\mathsf{packets}},\Delta^{\mathsf{tot}}_{\mathsf{packets}} \leftarrow$ Update\_Range($x^m, minP, maxP$)
	    \STATE $\xminpackets, \xmaxpackets  \leftarrow \updatestat (x^m, \Delta^{\mathsf{tot}}_{\mathsf{packets}}/s,  F^{\mathsf{packets}}_{i_{max}})$
		\STATE $\xsumbytes \leftarrow  \updateratio (x^m, \nabla^{\mathsf{tot}}_{\mathsf{bytes}}, F^{\mathsf{bytes}}_{i_{max}})$
		\STATE $\xsumbytes,\Delta^{\mathsf{tot}}_{\mathsf{bytes}}\leftarrow$ Update\_Range$(\xsumbytes, minB, maxB)$	
		\STATE $\xminbytes, \xmaxbytes  \leftarrow \updatestat (x^m, \Delta^{\mathsf{tot}}_{\mathsf{bytes}}/s,  F^{\mathsf{bytes}}_{i_{max}})$
	    \STATE$\xsumdur \leftarrow  \updateratio(x^m, \nabla^{\mathsf{tot}}_{\mathsf{dur}}, F^{\mathsf{dur}}_{i_{max}})$
		\STATE $\xsumdur, \Delta^{\mathsf{tot}}_{\mathsf{dur}} \leftarrow$ Update\_Range $(\xsumdur, minD, maxD)$
		\STATE $\xmindur, \xmaxdur  \leftarrow \updatestat ( x^m, \Delta^{\mathsf{tot}}_{\mathsf{dur}}/s,  F^{\mathsf{dur}}_{i_{max}})$
	\end{algorithmic}
\end{algorithm}
\vspace{-15pt}
		
\subsection{Malicious Domain Classifier}
\label{sec:madeapp}

The second threat detection application is to classify FQDN domain names contacted by enterprise hosts as \mal\ or \ben.  This security application has multiple types of feature-space constraints, including linear, non-linear, and statistical dependencies, and therefore can be used to test our \system\ framework for supporting multiple constraints.

\myparagraph{Dataset} We obtained access to a proprietary dataset
collected by a company that includes 89 domain features extracted from HTTP proxy logs collected at the border of an enterprise network.
This is the same dataset used for the design of MADE system for detecting malicious activity in enterprise networks and prioritizing the detected activities according to their risk described in detail by Oprea et al.~\cite{oprea2018made}. This dataset enables us to experiment with the variety of constraints in feature space, representative of security applications. Features are aggregated over multiple HTTP connections to the same external FQDN domain and are defined with the help of security experts. Each external FQDN is labeled as \mal\ or \ben. We group the modifiable features of MADE into a set of 7 families, included  in Table~\ref{tab:features_id}.  More details on all the features used in MADE are provided in the original paper~\cite{oprea2018made} (we preserved the feature ID for the MADE dataset in Table~\ref{tab:features_id}).

In this application, we do not have access to the raw HTTP traffic, only to features extracted from it and domain labels. Thus, the constraints are \textbf{mathematical constraints} in  feature space, for instance:

\begin{itemize}
\item For the Connection family, we have $NonLin$ dependence: computing average value over a number of events.
\item For the Bytes family, we need to update the ratio of two values and then update minimum, maximum and average values, thus, we have the combination of $Stat$ and $NonLin$ dependencies.
\item For the HTTP Method, we have the same  combination of $Stat$ and $NonLin$ dependencies as for Bytes family.
\item For the Content family, we need to ensure that the sum of all ratio values equals 1. This is a combination of $Lin$ and $NonLin$ dependencies.
\item For the Result Code, we also need to ensure that the sum of all fraction values equals to 1. Additionally, the number of connections with different codes must sum up to the total number of connections. This is a combination of $Lin$ and $NonLin$ dependencies.
\end{itemize}

\myparagraph{Attack algorithm} We assume that we add events to the logs, and never delete or modify existing events. For instance, we can insert more connections, as in the malicious connection classifier. 
The attack algorithm against the malicious domain classifier called the MADE attack follows the framework from Algorithm~\ref{alg:evasion_fw}. If the feature of the maximum gradient has no dependencies, it is just updated with the gradient value. Otherwise, every dependency family has a specific representative feature and is updated following one of the specified \updatedep\ functions. For example, for the Connection family, the representative feature is  \emph{Num\_Conn}, which is updated with the gradient value, and other features in this family are updated by calling the \updatestat\ function. The detailed algorithm for the \updatedep\ function of the MADE attack is illustrated in Algorithm~\ref{alg:made}. The functions for updating dependencies (e.g., \updatestat, \updatelin, \updatenonlin) are the same defined in the \system\ framework and discussed in Section~\ref{sec:dep}.

\begin{table}[H]
\centering
\begin{tabular}{|c||c||c||c|}
\hline
Family & Feature ID & Feature  & Description\\ 
\hline
Connections&1 & Num\_Conn&Number of established connections\\
&2& Avg\_Conn&Average number of connections per host\\
\hline
Bytes&3 & Total\_Recv\_Bytes&Total number of received bytes\\
&4&Total\_Sent\_Bytes&Total number of sent bytes\\
&5&Avg\_Ratio\_Bytes& Average ratio of received bytes\\ 
&&& over sent bytes per IP\\ 
&6&Min\_Ratio\_Bytes&Maximum ratio of received bytes\\ 
&&&over sent bytes per IP \\ 
&7&Max\_Ratio\_Bytes& Minimum ratio of received bytes\\
&&&over sent bytes per IP \\ 
\hline
HTTP & 8&Num\_POST & Total number of POST requests\\
Method &9&Num\_GET & Total number of GET requests\\
& 10 &Avg\_POST&Average number of POST requests\\
&&&over GET requests per IP\\
& 11 &Min\_POST&Minimum number of POST requests\\
&&&over GET requests per IP\\
& 12 &Max\_POST&Maximum number of POST requests\\
&&&over GET requests per IP\\
\hline
Content &46 &Frac\_empty & Fraction of connections with empty content type\\
&47&Frac\_js & Fraction of connections with js content type \\
&48&Frac\_html & Fraction of connections with html content type \\
&49&Frac\_img & Fraction of connections with image content type   \\
&50&Frac\_video & Fraction of connections with video content type   \\
&51&Frac\_text & Fraction of connections with text content type   \\
&52&Frac\_app & Fraction of connections with app content type   \\
\hline
Result & 59 & Num\_200 & Number of connections with result code 200\\
Code & 60 & Num\_300 & Number of connections with result code 300\\
&61&Num\_400& Number of connections with result code 400\\
&62&Num\_500&Number of connections with result code 500\\
&63&Frac\_200&Fraction of connections with result code 200\\
&64&Frac\_300&Fraction of connections with result code 300\\
&65&Frac\_400&Fraction of connections with result code 400\\
&66&Frac\_500&Fraction of connections with result code 500\\
\hline
Independent & 43 & Avg\_OS& Average number operating systems\\
&&&extracted from user-agent\\
& 44& Avg\_Browser& Average number of browsers used\\
&68&Dom\_Levels&Number of levels \\
&69&Sub\_Domains&Number of sub-domains \\
&70&Dom\_Length&Length of domain\\
&71&Reg\_Age& WHOIS registration age\\
&72&Reg\_Validity& WHOIS registration validity\\
&73&Update\_Age& WHOIS update age\\
&74&Update\_Validity&  WHOIS update validity\\
&75& Num\_ASNs & Number of ASNs\\
&76& Num\_Countries & Number of countries contacted the domain\\
\hline
\end{tabular}
\caption{Feature set for malicious domain classification that can be modified by the evasion attack.}
\label{tab:features_id}
\end{table}

\begin{algorithm}[H]
	\caption{MADE \updatedep\ ($s,x^m,\nabla,F_{i_{max}}$)}
	\label{alg:made}
	\begin{algorithmic}[1]
	\REQUIRE $s$: type of dependency family \\
	$x^m$: data point in iteration $m$\\
	$\nabla$: gradient of objective with respect to $x$ \\
	$F_{i_{max}}$: all dependent features on $i_{max}$\\
	$ i_{max}$: feature of maximum gradient
		\IF{$s == \mbox{Stat}$}
		    \STATE $v \leftarrow $NEW\_VALUE\_ADDED$(x^m, \nabla, F_{i_{max}})$
			\STATE $\updatestat(x^m, v, F_{i_{max}})$
		\ENDIF
		\IF{$s == \mbox{Lin}$}
			\STATE $\updatelin(x^m,\nabla,F_{i_{max}})$
		\ENDIF	
		\IF{$s == \mbox{NonLin}$}
			\STATE $\updatenonlin(x^m,\nabla,F_{i_{max}})$
		\ENDIF
		\IF {$s == \mbox{Combination}$}
			\STATE $\updatecomb(x^m,\nabla,F_{i_{max}})$
		\ENDIF	
	\end{algorithmic}
\end{algorithm}
\section{Experimental evaluation for network traffic classifier}
\label{sec:neriseval}

We evaluate \system\ for the malicious network traffic classifier trained with both the raw data and engineered feature representations. We show feasible attacks that insert a small number of network connections to change the \mal\ prediction to \ben. We only analyze the \system\ attack with the Projected optimization objective here. In the following section, we  analyze our \system\ framework for the malicious domain classifier for both the Projected and Penalty attacks. 

\subsection{Experimental setup}
\label{sec:ctu_dataset}

CTU-13 is a collection of 13  scenarios including both legitimate traffic from a university campus network, as well as labeled connections of malicious botnets~\cite{Garca2014AnEC}. We restrict to three scenarios for the Neris botnet (1, 2, and 9). We choose to train on two of the scenarios and test the models on the third, to guarantee independence between training and testing data.  

The raw data representation has 3,712,935 data points,  from which 151,625 are labeled as botnets. The attacker can modify three features per connection: bytes and packets sent, and duration. The training data in the engineered features representation has 3869 \mal\ examples, and 194,259 \ben\ examples, and an imbalance ratio of 1:50. There is a set of 432 statistical features that the attacker can modify (the ones that correspond to the characteristics of sent traffic on 17 ports). 
The physical constraints and statistical dependencies in both scenarios have been detailed in Section~\ref{sec:nerisattack}. 

We considered two baseline attacks: \emph{Baseline 1} (in which the features that are modified iteratively are selected at random), and \emph{Baseline 2} (in which, additionally, the amount of perturbation is sampled from a standard normal distribution $N(0,1)$).

\subsection{Attack results for raw data representation}
\label{sec:neris_attack_raw} 

For training we have used \ffnn\ with two layers and a sigmoid activation function. The architecture that corresponds to the best performance has 12 neurons in the first layer, and 1 neuron in the second layer. We have trained it using Adam optimizer with a learning rate equal to 0.0001 for 20 epochs with batch size 64. The best results are for training on scenarios 2 and 9, and testing on scenario 1, with an F1 score of 0.70.

We consider an attack on testing scenario 1, and the success rate of our attack is 100\% already at a small $L_2$ distance of 2. Intuitively, an attacker can add a few packets and bytes to a connection and change its classification easily. We  compare its performance to  Baseline 2, which achieves only 73\% success rate at $L_2$ distance of~2.

\subsection{Attack results for engineered features}
\label{sec:neris_attack}

\begin{figure*}[t]
\centering
  \begin{subfigure}[b]{0.3\linewidth}
    \includegraphics[width=\linewidth]{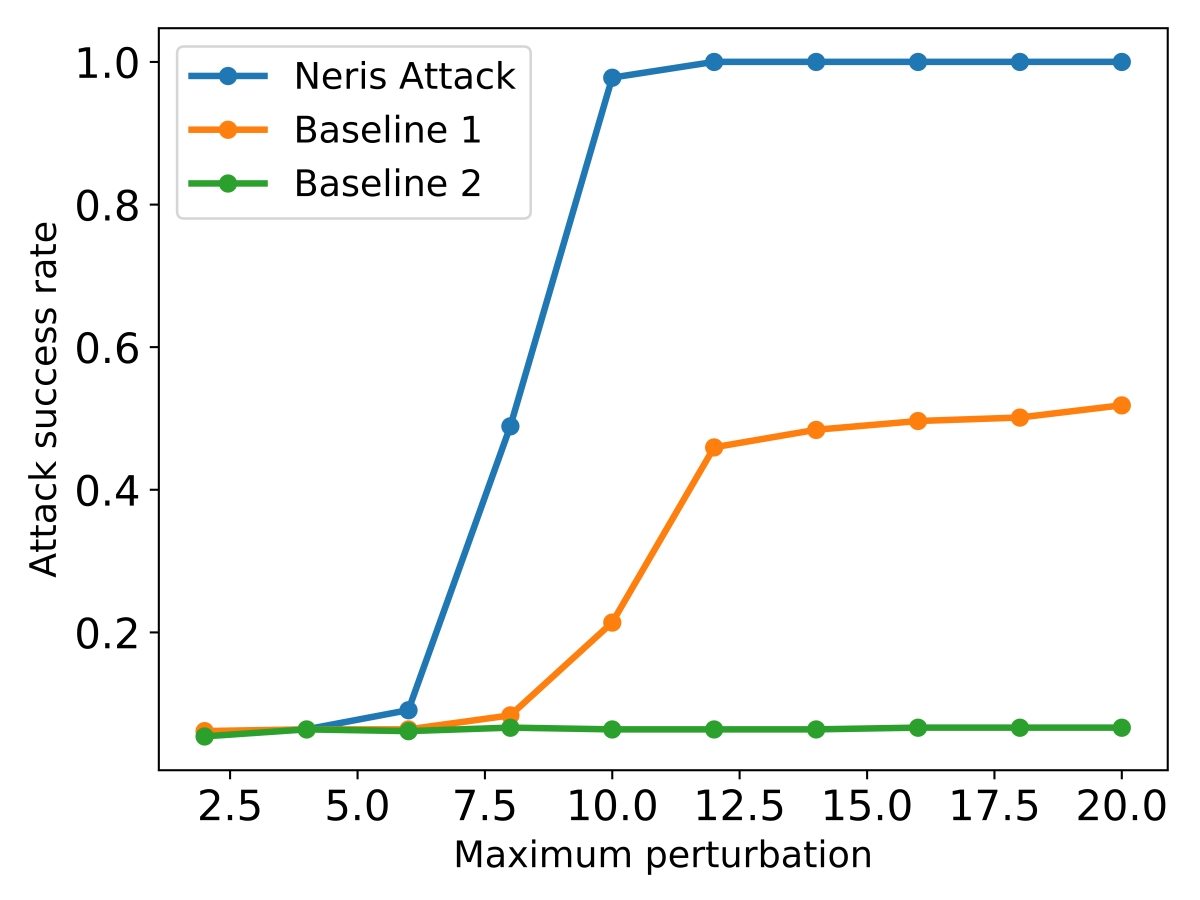}
    \caption{ FENCE Projected attack\\ success rate.}
    \label{fig:neris_sr}
  \end{subfigure}
  \begin{subfigure}[b]{0.3\linewidth}
    \includegraphics[width=\linewidth]{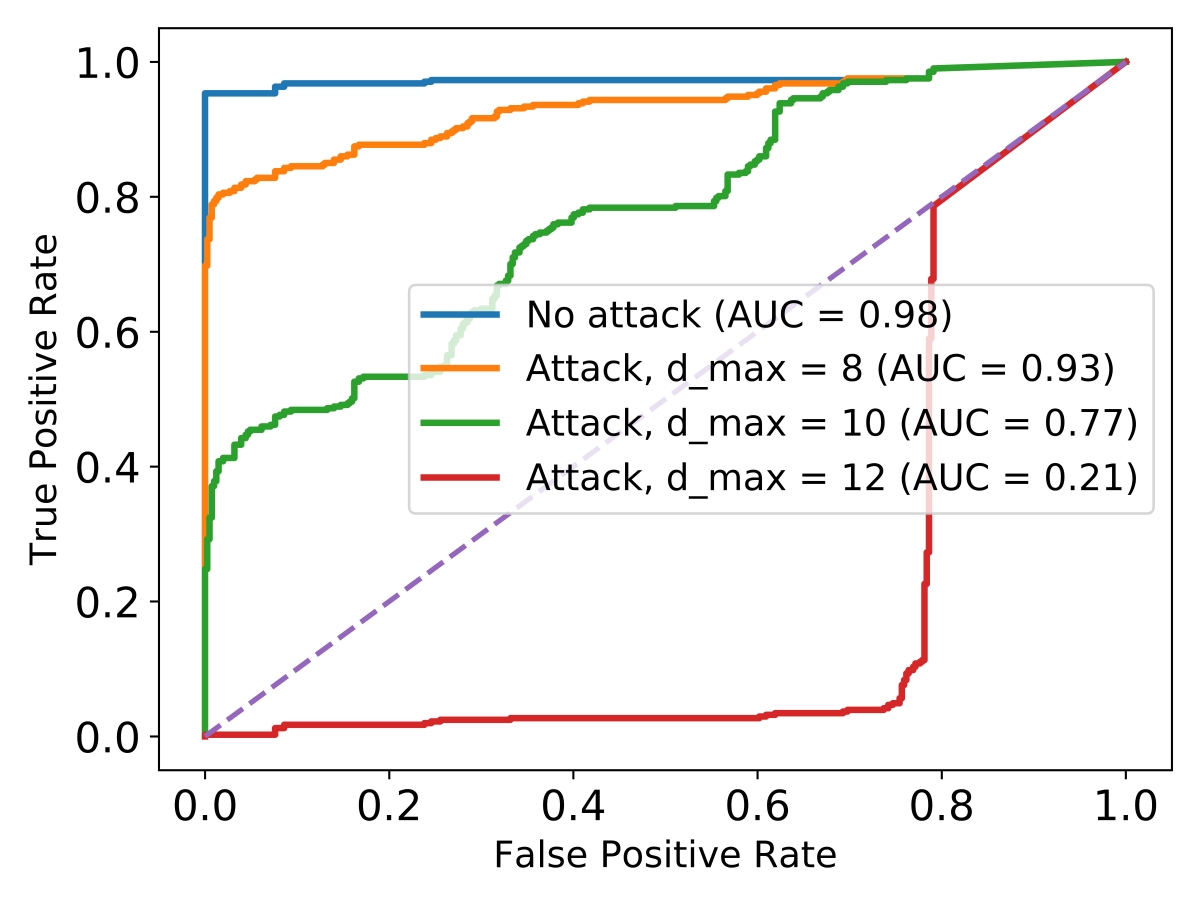}
    \caption{ROC curves under FENCE\\ Projected attack.}
    \label{fig:neris_rocs}
  \end{subfigure}
  \begin{subfigure}[b]{0.3\linewidth}
    \includegraphics[width=\linewidth]{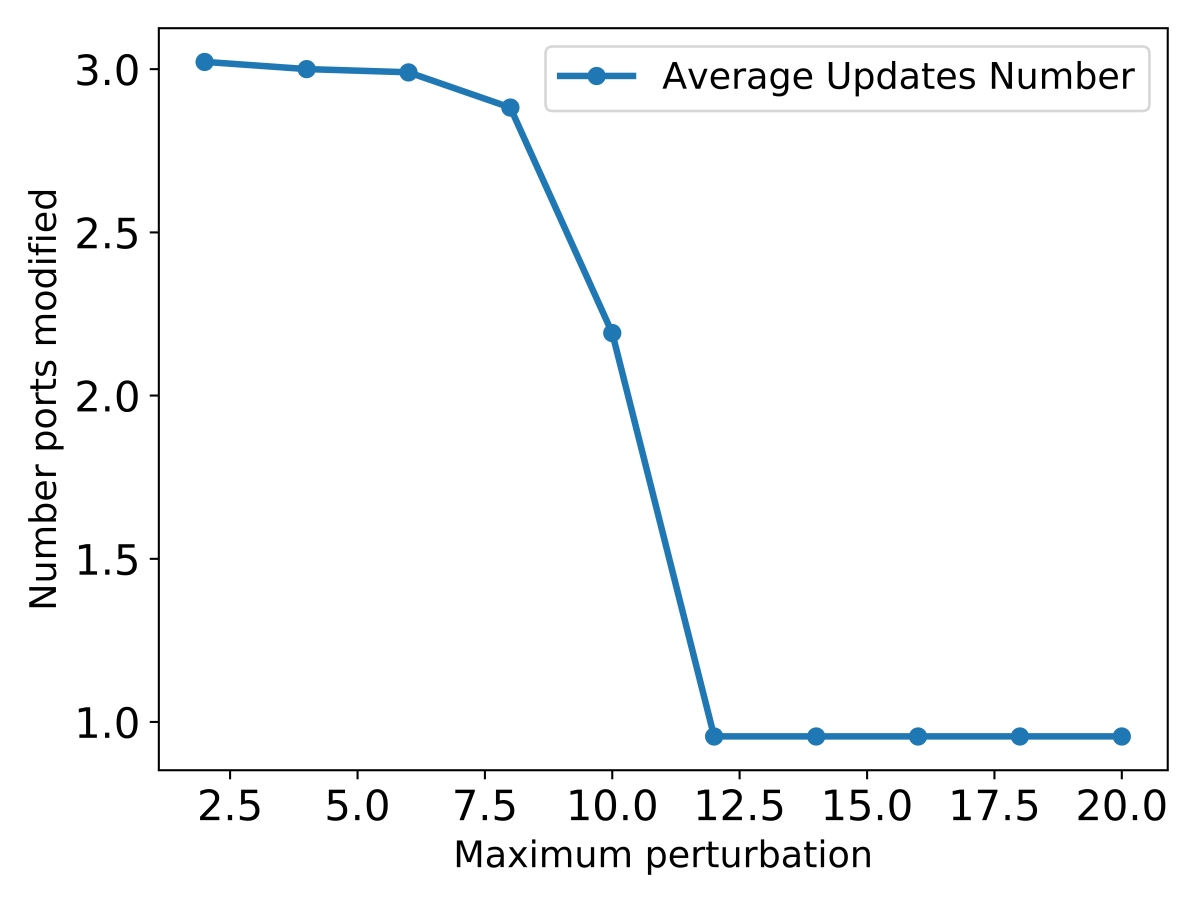}
    \caption{Average number of updated\\ ports.}
    \label{fig:neris_ports_ave}
 \end{subfigure}
\caption{Projected attack results on network traffic classifier.}
\end{figure*}

We perform model selection and training for a number of \ffnn\ architectures on all combinations of two scenarios, and tested the models for generality on the third scenario. The best architecture consists of three layers with 256, 128 and 64 hidden layers. We used the Adam optimizer, 50 epochs for training, a mini-batch of 64, and a learning rate of 0.00026. The F1 and AUC scores are much better than the \ffnn\ based on raw data representation. For instance, the best scenario is training on 1, 9, and testing on 2, which achieve an F1 score of 0.97, compared to 0.70 for raw data.

We thus perform a more extensive analysis of the attack against engineered features in this scenario. The testing data for the attack is 407 \mal\ examples from scenario 2, among which 397 were predicted correctly by the classifier.

\myparagraph{Evasion attack performance} First, we analyze the attack success rate with respect to the  allowed perturbation, shown in Figure~\ref{fig:neris_sr}. The attack reaches 99\% success rate at  $L_2$ distance 16. Interestingly, in this case the two baselines perform poorly, demonstrating again the clear advantages of our framework. We plot next the ROC curves under evasion attack in Figure~\ref{fig:neris_rocs} (using the 407 \mal\ examples and 407 \ben\ examples from testing scenario 2). At distance 8, the AUC score is 0.93 (compared to 0.98 without adversarial examples), but there is a sudden change at distance 10, with the AUC score dropping to 0.77. Moreover, at distance 12, the AUC reaches 0.12, showing the model's degradation under evasion attack with relatively small distance.

\begin{table*}[t]
\centering
\begin{tabular}{|l||l||l||l||l||l||l||l||l|}
\hline
ts &port & prot & duration &  o\_bts & r\_bts & o\_pkts & r\_pkts & state\\ 
\hline
1 &53&	UDP & 2.26638&67 &	558&2 &	2&SF\\
2 &	13363&	TCP	& 444.334 &	707&	671&	14	&11& SF\\
3 &	 1035 & TCP	&	276.084218 &	20768&	0&	110 &0& OTH\\
\hline

4 &443  & TCP & \color{red}432.47 &\color{red}112404& 0 & \color{red}87  &0  &  OTH \\

\hline
\end{tabular}
\caption{Example of  Zeek logs records (top 3 rows), and one of the 12 attacker-generated connection logs added for the adversarial example perturbation (4th row).}
\label{tab:Brologs}
\end{table*}

\begin{table}[ht]
\centering
\begin{tabular}{|c||c||c||c||c||c||c||c|}
\hline
Feature & Average & Std. Dev. & 25 \%& 50 \%& 75 \% & 95 \% & Maximum \\ 
\hline
Duration &552.20&7872.84&0.13&5.71&79.38&1578.82&1060383.73\\
Orig\_Bytes &150756.59 & 12976082.76 &96&415&2466&51117.3&2091734081\\
Orig\_Pkts &342.75 & 5221.95 &2&9&34&522&405463\\
\hline
\end{tabular}
\caption{Feature statistics in training data.}
\label{tab:neris_features_stats}
\end{table}

\begin{table}[ht]
\centering
\begin{tabular}{|c|c||c||c||c||c||c||c|}
\hline

Setting & Average & Std. Dev. & 25 \%& 50 \%& 75 \% & 95 \% & Maximum \\ 
\hline
Training data & 38.12&98.85&2.42&4.52&24.53&195.58&890.56\\
Projected perturbation & 7.66&0.57&7.66&7.85&7.93&7.97&7.99\\
\hline
\end{tabular}
\caption{$L_2$ norm statistics for training data (top row) and perturbation added by the Projected attack (bottom row).}
\label{tab:l2_input_stats}
\end{table}

The average number of port families updated during the attack is shown in Figure~\ref{fig:neris_ports_ave}. The maximum number is 3 ports, but it decreases to 1 port at distance higher than 12. While counter-intuitive,  at larger distances the attacker can add larger perturbation to the aggregated statistics of one port, crossing the decision boundary. The ports most frequently modified are 443 and 80, which are the ports with most network traffic.

\myparagraph{Adversarial examples} We show an adversarial example generated by the Projected attack at distance 14. 
The attacker adds only 12 TCP connections on port 443,  including 87 packets, each of size 1292 bytes, with connection duration of 432.47 seconds.  Table~\ref{tab:Brologs} shows  one of the 12 attacker-generated connections to create the adversarial example. The destination IP can be selected by the attacker so that it is under its control and does not send any bytes or packets.  These new connections are added to the activity the attacker already does inside the network, so the malicious functionality of the attack is preserved. Interestingly, all adversarial attacks succeed with at most 12 new connections at distances higher than 10. 
In Table~\ref{tab:neris_features_stats} we show  statistics for the duration, sent bytes and sent packets features in the training data. We make the observation that the feature values in the resulting adversarial example are below the average feature values  of the training data. In particular, the Orig\_Bytes feature has an average value 150KB and a very high standard deviation (12.9MB), while the adversarial example only uses 112,404 bytes, which is very little communication (112KB).

Furthermore, we illustrate the $L_2$ norm statistics of samples in the training data along with $L_2$ norms of perturbations   added to create adversarial examples at distance 8 in Table~\ref{tab:l2_input_stats}. We observe that the average $L_2$ perturbation norm (7.66) is  4.97 times lower than the training data average $L_2$ norm (38.12). Given that the standard deviation (0.57) and maximum value (7.99) of the  perturbation $L_2$ norm are small compared to the standard deviation (98.85) and  the maximum value  (890.56) of the $L_2$ norm of training samples,  we consider the resulting adversarial attacks stealthy.

\section{Experimental evaluation for malicious domain classifier}
\label{sec:madeeval}

In this section we perform a detailed evaluation of the \system\ attack on the MADE malicious domain classifier trained on the enterprise dataset~\cite{oprea2018made}. We compare the Projected and Penalty optimization methods and analyze the impact of imbalanced training datasets. We also test the transferability of the \system\ attacks across other models and  architectures and evaluate the potential of adversarial training as mitigation against \system\ evasion attacks.

\subsection{Experimental setup}
\label{sec:dataset}

The data for training and testing the models was extracted from security logs  collected by web proxies at the border of a large enterprise network with over 100,000 hosts.  The number of monitored external domains in the training set is 227,033, among which 1730 are classified as \mal\ and 225,303 are \ben. For training, we sampled a subset of training data to include 1230 \mal\ domains, and a different number of \ben\ domains to get several imbalance ratios between the two classes (1, 5, 15, 25, and 50). We used the remaining 500 \mal\ domains and sampled 500 \ben\ domains for testing the evasion attack. Overall, the dataset includes 89 features from 7 categories. 

Among the features included in the dataset, we determined a set of 31 features that can be modified by an attacker (see Table~\ref{tab:features_id} for their description). These include communication-related features (e.g., number of connections, number of bytes sent and received, etc.), as well as some independent features (e.g., number of levels in the domain or domain registration age). Other features in the dataset (for example, those using URL parameters or values) are more difficult to change, and we consider them immutable during the evasion attack.

This dataset is extremely imbalanced, and we sample a different number of \ben\ domains from the data, to control the imbalance ratio. We are interested in how the imbalance affects the attack success rate. On this dataset, we also compare the Projected and Penalty attack objectives.

\subsection{\system\ attack evaluation}
\label{sec:exp_adv}

We experimented with several models for training classifiers, including logistic regression, random forest, and different \ffnn\ architectures. The best performance was achieved by a two-layer \ffnn\ with 80 neurons in the first layer, and 50 neurons in the second layer. ReLU activation function is used after all hidden layers except for the last layer, which uses sigmoid. We used the Adam optimizer and SGD with different learning rates. The best results were obtained with Adam and a learning rate of 0.0003. We trained for 75 epochs with a mini-batch size of 32. The resulting model had an AUC score of 89\% with cross-validation, in the balanced case. These results were comparable to the best random forest model we trained and better than logistic regression.

\begin{figure*}[ht]
\centering
  \begin{subfigure}[b]{0.4\linewidth}
    \includegraphics[width=\linewidth]{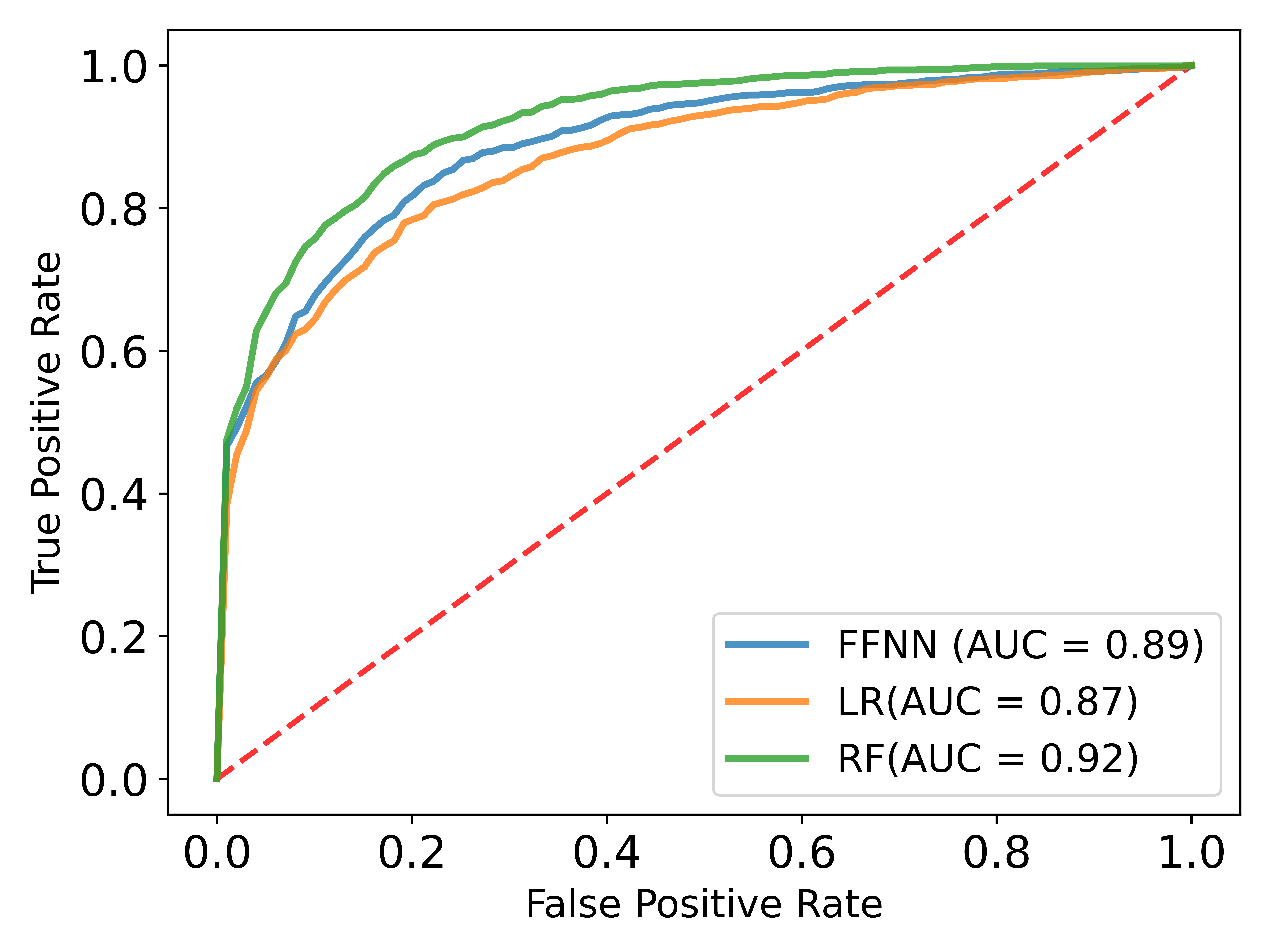}
    \caption{Models performance.}
    \label{fig:models}
  \end{subfigure}
  \begin{subfigure}[b]{0.4\linewidth}
    \includegraphics[width=\linewidth]{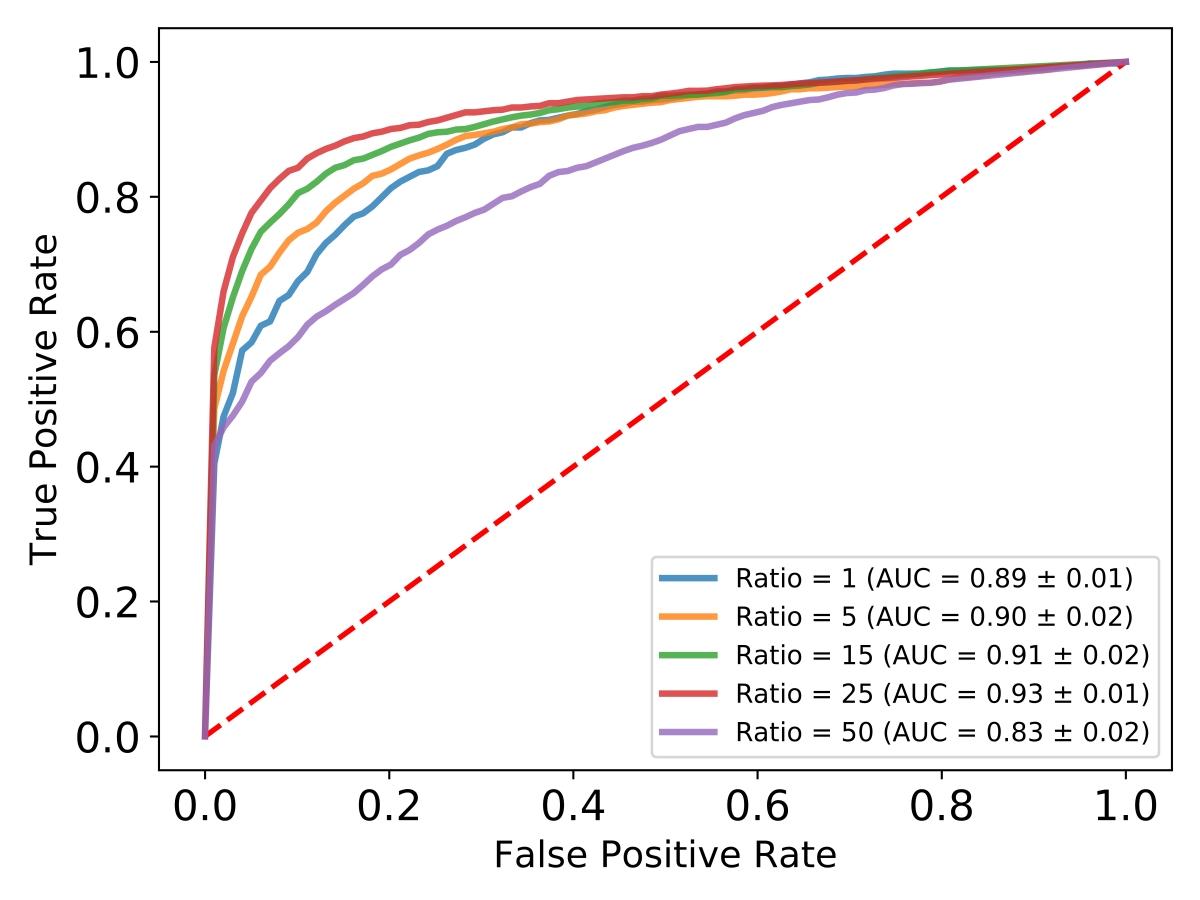}
    \caption{Imbalance results on \ffnn.}
  \end{subfigure}
\caption{Training results for malicious domain classifier.}
\label{fig:ffnn_r}
\end{figure*}

The ROC curves for training logistic regression, random forest and \ffnn\ are given in Figure~\ref{fig:ffnn_r} (a), while the results for \ffnn\ with different imbalanced ratios are in Figure~\ref{fig:ffnn_r} (b). Interestingly, the performance of the model increases to 93\% AUC for an imbalance ratio up to 25, after which it starts to decrease (with AUC of 83\% at a ratio of 50). Our intuition is that the \ffnn\ model achieves better performance when more training data is available (up to a ratio of 25). But once the \ben\ class dominates the \mal\ one (at ratio of 50), the model performance starts to degrade.

\begin{figure*}[t]
\centering
  \begin{subfigure}[b]{0.32\linewidth}
    \includegraphics[width=\linewidth]{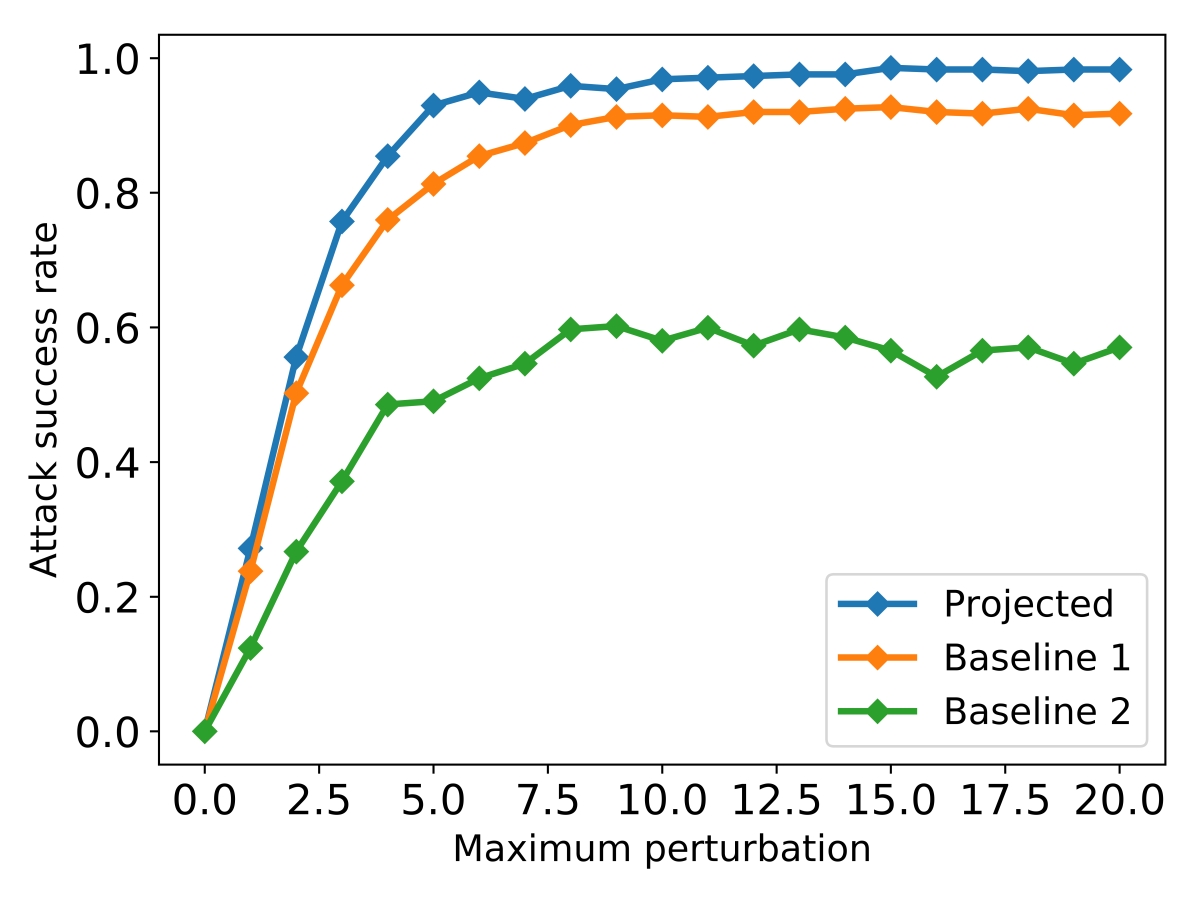}
    \caption{FENCE attack success rate.}
    \label{fig:proj_rand}
  \end{subfigure}
  \begin{subfigure}[b]{0.32\linewidth}
    \includegraphics[width=\linewidth]{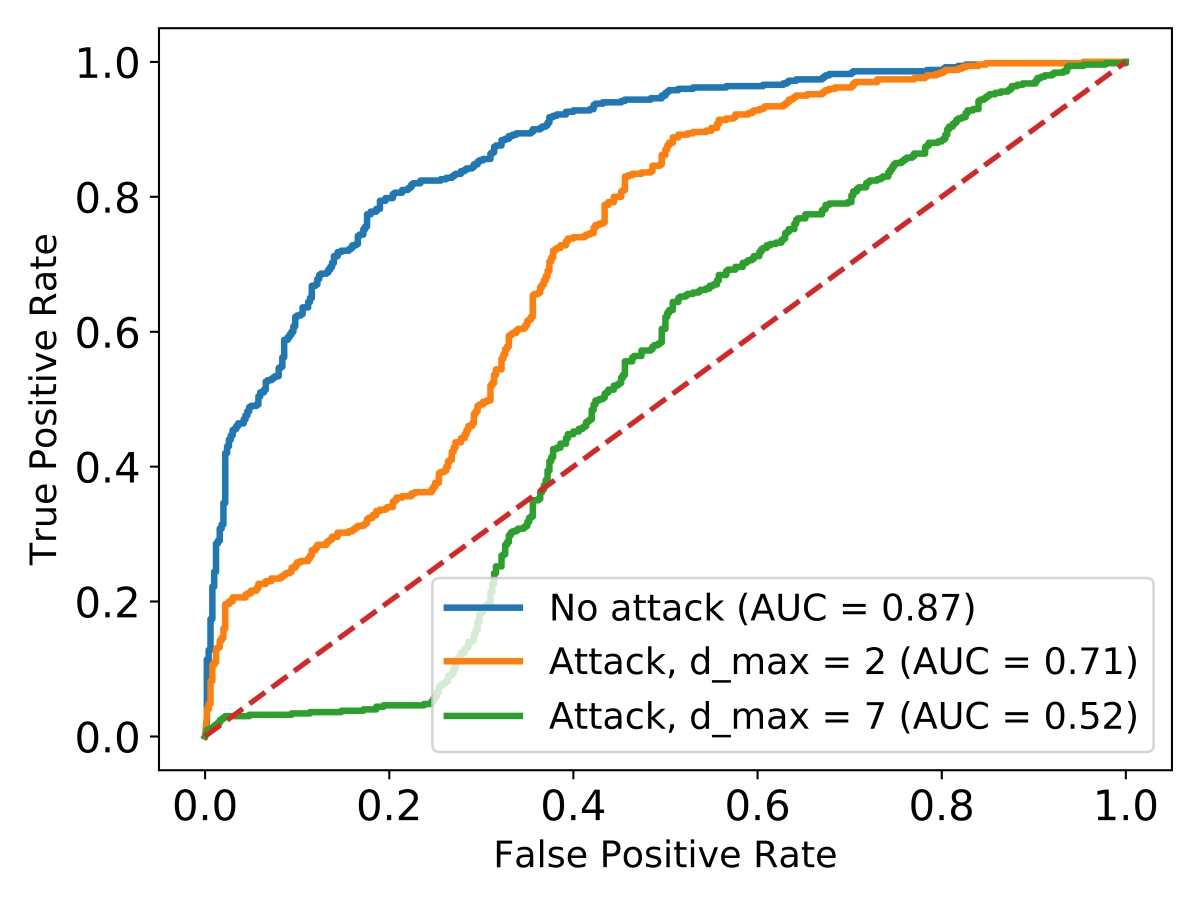}
    \caption{ROC curves under attack.}
    \label{fig:proj_roc}
  \end{subfigure}
  \begin{subfigure}[b]{0.32\linewidth}
    \includegraphics[width=\linewidth]{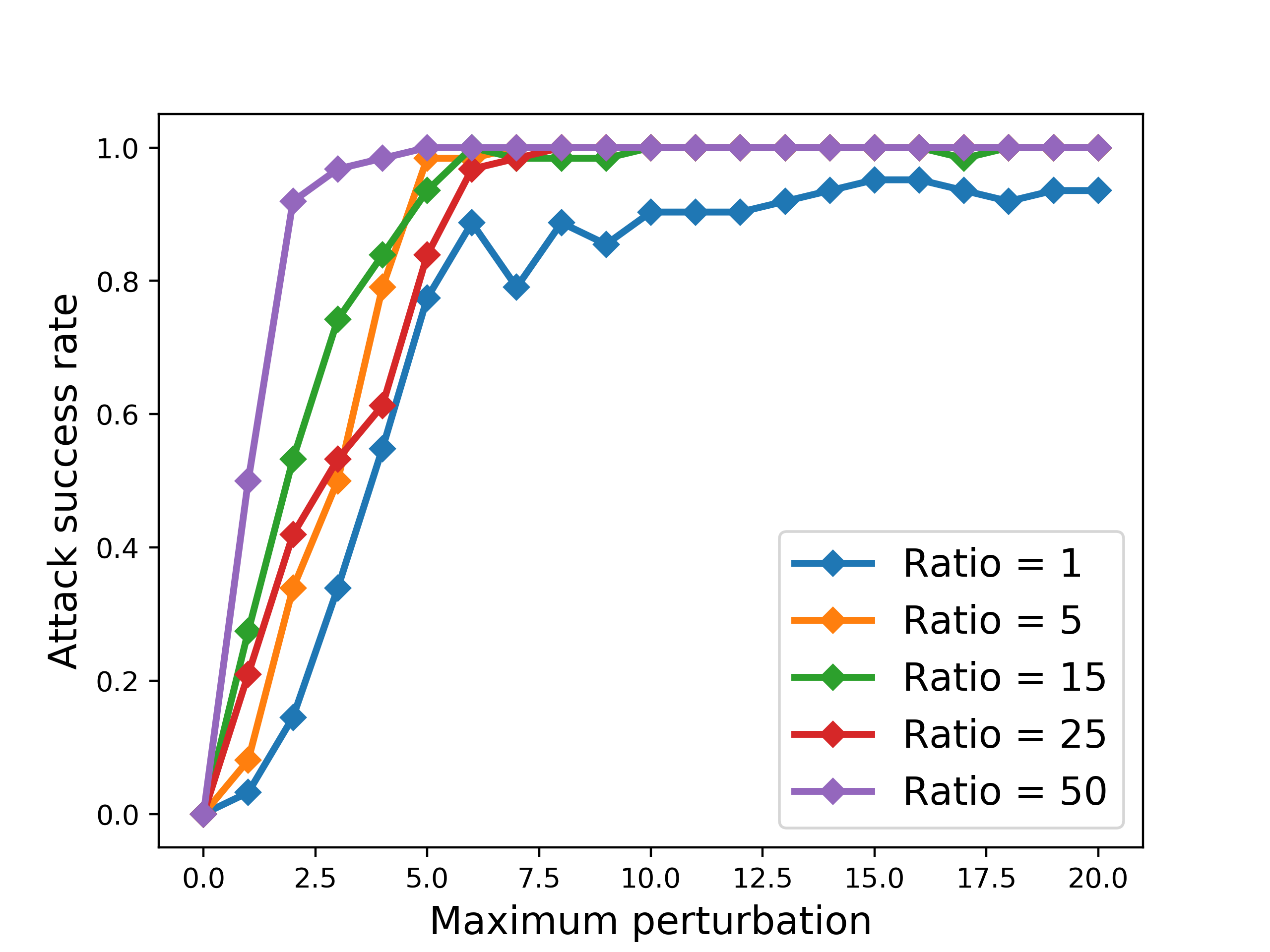}
    \caption{Imbalance sensitivity.}
    \label{fig:proj_mal}
  \end{subfigure}
\caption{\system\ Projected attack results for the malicious domain classifier. }
\end{figure*}

\myparagraph{\system\ Projected attack results} We evaluate the success rate of the attack with Projected objective first for balanced classes (1:1 ratio). We compare in Figure~\ref{fig:proj_rand} the attack against the two baselines. The attacks are run on 412 \mal\ testing examples classified correctly by the \ffnn.  The Projected attack improves both baselines, with Baseline 2 performing much worse, reaching success rate of 57\% at a distance of 20, and Baseline 1 has a success of 91.7\% compared to our attack (98.3\% success). This shows that the attacks are still performing reasonably if feature selection is done randomly, but it is very important to add perturbation to features consistent with the optimization objective.

We also measure in Figure~\ref{fig:proj_roc} the decrease in the model’s performance before and after the evasion attack at different perturbations (using 500 \mal\ and 500 \ben\ examples not used in training). While the AUC score is 0.87 originally, it drastically decreases to 0.52 under evasion attack at perturbation 7. This shows the significant degradation of the model’s performance under evasion attack.

Finally, we run the attack at different imbalance ratios and measure its success for different perturbations. In this experiment, we select 62 test examples that all models (trained for different imbalance ratios) classified correctly before the evasion attack.  The results are illustrated in Figure~\ref{fig:proj_mal}. At $L_2$ distance 20, the evasion attack achieves a 100\% success rate for all ratios except 1. 
Additionally, we observe that with a higher imbalance, it is easier for the attacker to find  adversarial examples (at a fixed distance). One reason is that models that have lower performance (such as the model trained with 1:50) are easier to attack. Second, we believe that as the imbalance gets higher the model becomes more biased towards the majority class (\ben), which is the target class of the attacker, making it easier to cross the decision boundary between classes.

We include an adversarial example in Table~\ref{tab:proj_adv}. We only show the features that are modified by the attack and their original value. As we observe, the attack preserves the feature dependencies: the average ratio of received bytes over sent bytes (Avg\_Ratio\_Bytes) is consistent with the number of received (Total\_Recv\_Bytes) and sent (Total\_Sent\_Bytes) bytes. In addition, the attack modifies the domain registration age, an independent feature, relevant to malicious domain classification~\cite{MaKDD09}. However, there is a higher cost to change this feature: the attacker should register a malicious domain and wait to get a larger registration age. If this cost is prohibitive, we can easily modify our framework to make this feature immutable.

\begin{table}[ht]
\centering
\begin{tabular}{|c||c||c|}
\hline
Feature & Original & Adversarial  \\ 
\hline
NIP &1 & 1\\
Total\_Recv\_Bytes & 32.32 &  \color{red}{43653.50}\\
Total\_Sent\_Bytes & 2.0 & \color{red}2702.62\\
Avg\_Ratio\_Bytes & 16.15& 16.15\\
Registration\_Age & 349&\color{red}3616\\
\hline
\end{tabular}
\caption{Adversarial example for the \system\ Projected attack at distance 10.}
\label{tab:proj_adv}
\end{table}

 
We constructed 45 adversarial examples at $L_2$ distance 20 and calculated the average perturbation for every feature that was modified to show that the adversarial examples generated at this distance are unnoticeable.  The results can be found in Table~\ref{tab:proj_ave_features}. Additionally, we include  statistics for the  features on the training dataset in Table~\ref{tab:features_stats_made}. We observe that the generated adversarial examples have stealthy perturbations, given the feature distribution and meaning.  For example, the number of sent bytes is increased by 17195.3 (17.19KB), the registration age of a domain is increased on average by 1.14 days, while the number of sub domains is increased by 0.28 on average.Moreover, all average perturbations are much smaller than the standard deviation of features in the training data. For instance, the average perturbation of the sub domains number is 0.28 compared to the corresponding   standard deviation of 2867.53 in the training data. This additionally confirms the fact that the generated perturbations for adversarial examples can be considered unnoticeable.

Finally, we show the $L_2$ norm statistics of samples in the training data along with the $L_2$ norm of perturbations added to create adversarial examples by the Projected attack at distance 5 in the first two rows of Table~\ref{tab:l2_perturb_stats_made}. Even if we used a distance of 5 in the Projected attack, many of the generated perturbations have lower norm (in particular, half of the adversarial examples at distance 5 have norm lower than 3.17). The average $L_2$ norm of perturbation (3.20) is much smaller than the standard deviation of the training samples $L_2$ norm (26.9), resulting in relatively stealthy perturbations.  Additionally, the maximum $L_2$ norm of perturbation (5) is only a 0.005 fraction of the maximum possible $L_2$ norm of the training samples (890).

\begin{table}[ht]
\centering
\begin{tabular}{|c||c|}
\hline
Feature & Average perturbation  \\ 
\hline
Total\_Recv\_Bytes & 96340\\
Total\_Sent\_Bytes & 17195.3\\
Min\_Ratio\_Bytes & 23.26 \\
Sub\_Domains & 0.28\\
Reg\_Age & 1.14\\
Reg\_Validity & 158.91\\
Update\_Age & 24.98\\
Update\_Validity & 59.53\\
\hline
\end{tabular}
\caption{Average modified features perturbation for the \system\ Projected attack at distance 20.}
\label{tab:proj_ave_features}
\end{table}

\begin{table}[ht]
\centering
\begin{tabular}{|c||c||c||c||c||c||c||c|}
\hline
Feature & Average & Std. Dev. & 25 \%& 50 \%& 75 \% & 95 \% & Maximum \\ 
\hline
Total\_Recv\_Bytes &3332.4 &84148.49&96.09&433.17&1420.07&7092.40&24854191.98\\
Total\_Sent\_Bytes &213.76 & 52373.22 &6.16&14.54&31.48&120.46&25133008.74\\
Min\_Ratio\_Bytes &94.56 & 3298.47 &2.43&17.58&48.42&181.73&1190540.7\\
Num\_GET&77.22&1704.09&12&30&62&198&746776\\
Sub\_Domains & 485.66 & 2867.53&1&1&2&1247&54632\\
Reg\_Age & 2377.44&1880.03&820&2404&3179&6134&16649\\
Reg\_Validity & 2914.57 &2158.44&1096&2927&4017&6940&37114\\
Update\_Age & 348.68 &486.29&96&295&385&1128&42215\\
Update\_Validity & 839.28 &888.90&365&605&908&2928&43587\\
\hline
\end{tabular}
\caption{Feature statistics in training data.}
\label{tab:features_stats_made}
\end{table}

\begin{table}[ht]
\centering
\begin{tabular}{|c|c||c||c||c||c||c||c|}
\hline

Setting & Average & Std. Dev. & 25 \%& 50 \%& 75 \% & 95 \% & Maximum \\ 
\hline
Training data & 5.79&26.9&3.89&4.52&5.63&10.4&890\\
Projected perturbation & 3.20&1.20&2.17&3.17&4.21&4.99&5\\
Penalty perturbation & 2.76&1.03&1.92&2.56&3.70&4.44& 4.92\\

\hline
\end{tabular}
\caption{$L_2$ norm statistics for feature vectors in training data (first row), perturbation added by the Projected attack   (second row, 89\% success rate), and perturbation for the Penalty attack  (third row, 70\% success rate) at distance 5.}
\label{tab:l2_perturb_stats_made}
\end{table}

\begin{figure*}[ht]
\centering
  \begin{subfigure}[b]{0.3\textwidth}
    \includegraphics[width=\textwidth]{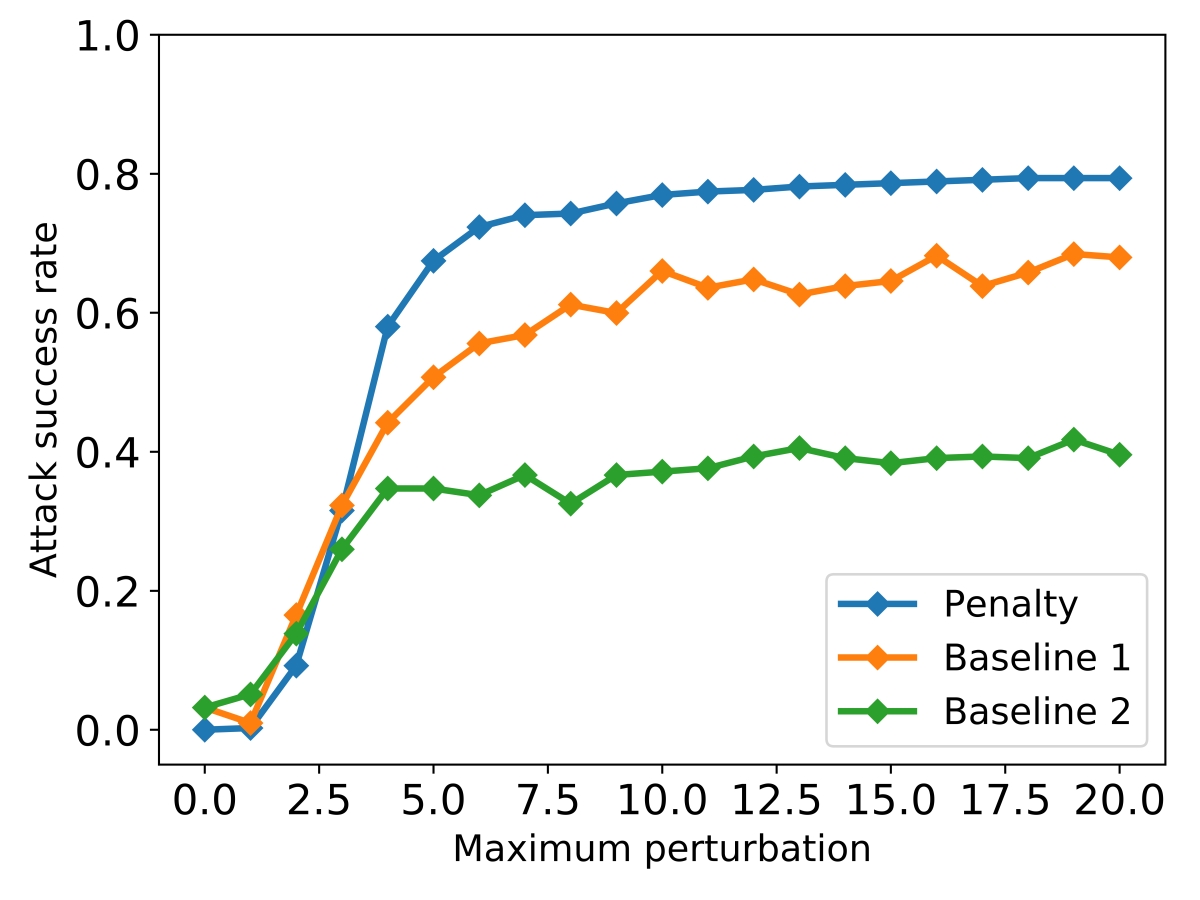}
    \caption{FENCE attack success rate.}
    \label{fig:pen_rand}
  \end{subfigure}
  \begin{subfigure}[b]{0.3\textwidth}
    \includegraphics[width=\textwidth]{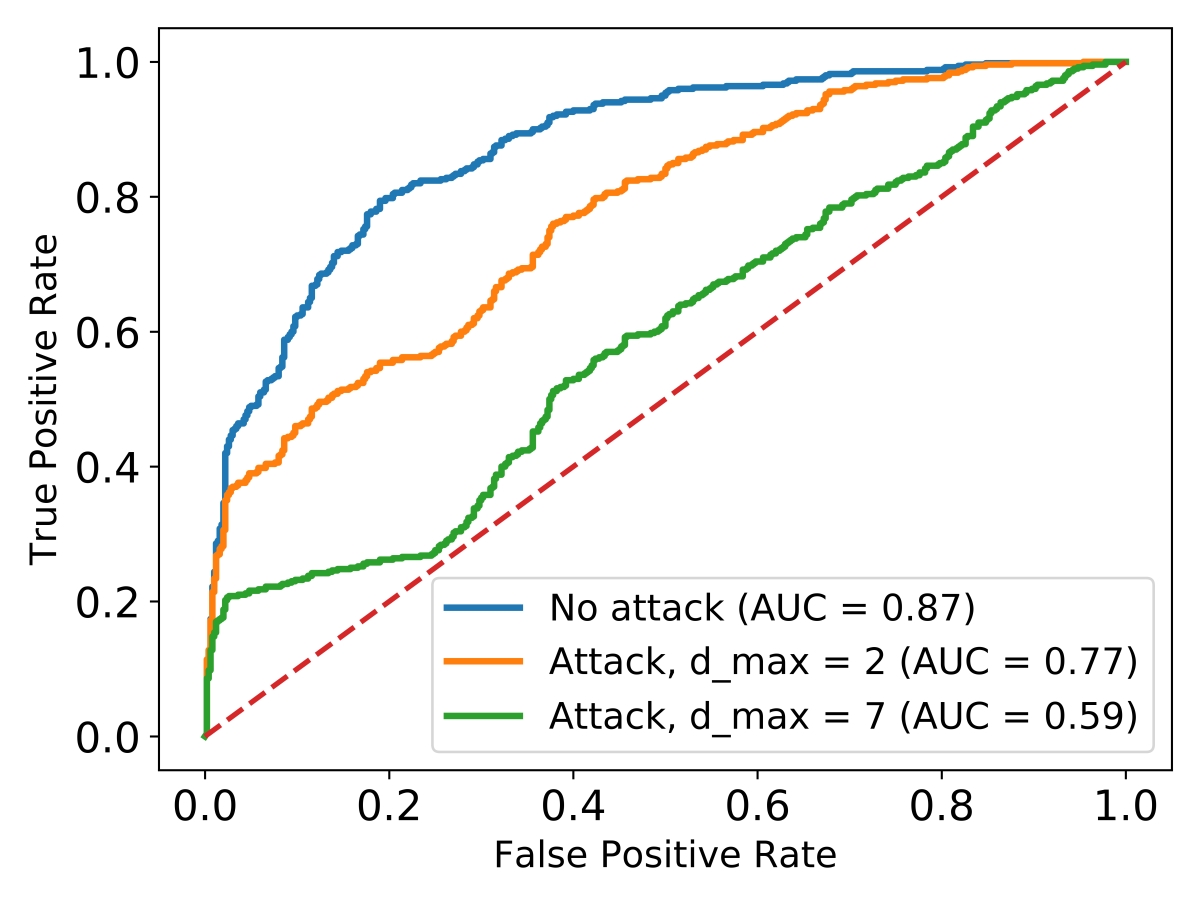}
    \caption{ROC curves under attack.}
    \label{fig:pen_roc}
  \end{subfigure}
  \begin{subfigure}[b]{0.3\textwidth}
    \includegraphics[width=\textwidth]{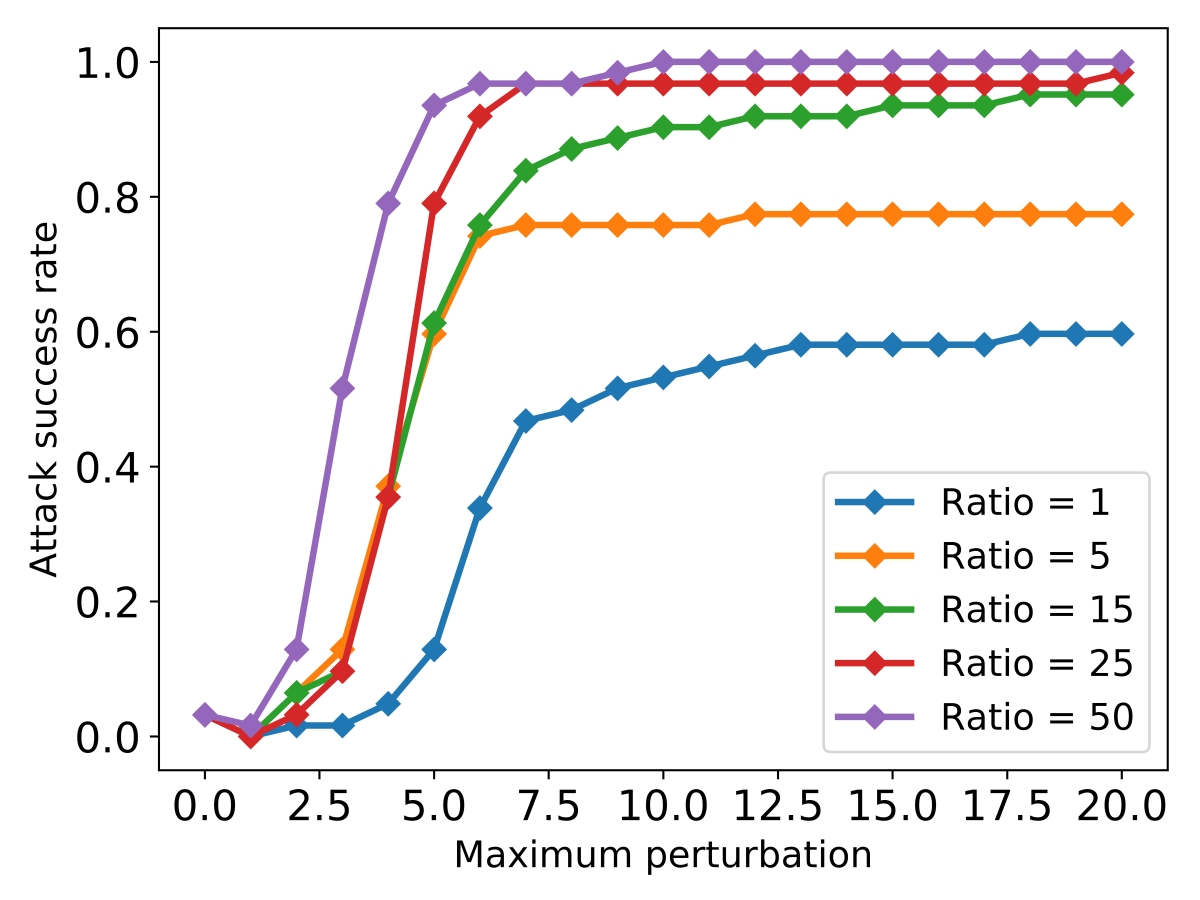}
    \caption{Imbalance sensitivity.}
    \label{fig:pen_mal}
  \end{subfigure}
\caption{FENCE Penalty attack results for malicious domain classifier.}
\end{figure*}

\myparagraph{\system\ Penalty attack results}
We now discuss the results achieved by applying our attack with the Penalty objective on the testing examples. Similar to the Projected attack, we compare the success rate of the Penalty attack to the two types of baseline attacks for balanced classes, in Figure~\ref{fig:pen_rand} (using the 412 \mal\ testing examples classified correctly). Overall, the Penalty objective is performing worse than the Projected one, reaching 79\% success rate at $L_2$ distance of 20. We observe that in this case both baselines perform worse, and the attack improves upon both baselines significantly. The decrease of the model’s performance under the Penalty attack is illustrated in Figure~\ref{fig:pen_roc} (for 500 \mal\ and 500 \ben\ testing examples). While AUC is 0.87 originally on the testing dataset, it decreases to 0.59 under the evasion attacks at the maximum allowed perturbation of 7. Furthermore, we measure the attack success rate at different imbalance ratios in Figure~\ref{fig:pen_mal} (using the 62 testing examples classified correctly by all models). For each ratio value we  searched for the best hyper-parameter $c$ between 0 and 1 with step 0.05. Here, as with the Projected attack, we see the same trend: as the imbalance ratio gets higher, the attack performs better, and it works best at imbalance ratio of 50.

We include an adversarial example generated by the Penalty attack in Table~\ref{tab:pen_adv}. We only show the features that were updated by the attack, which modifies only the amount of bytes sent (by 4.7KB) and the bytes received (by 73KB). The attack is able to preserve the $Ratio$ dependency: Avg\_Ratio\_Bytes = Total\_Recv\_Bytes/Total\_Sent\_Bytes/NIP.

\begin{table}[ht]
\centering
\begin{tabular}{|c||c||c|}
\hline
Feature & Original & Adversarial  \\ 
\hline
NIP &1 & 1\\
Total\_Recv\_Bytes &146.77&\color{red}73016.06\\
Total\_Sent\_Bytes&9.55&\color{red}4752.92\\
Avg\_Ratio\_Bytes&15.36&15.36\\
\hline
\end{tabular}
\caption{Adversarial example for the \system\ Penalty attack at distance 10.}
\label{tab:pen_adv}
\end{table}

We constructed 45 adversarial examples at $L_2$ distance 20 for the Penalty attack and calculated the average perturbation for every  modified feature. The results are illustrated in Table~\ref{tab:pen_ave_features}. Given the meaning of the features, we can conclude that an $L_2$ distance  of 20 can be considered reasonable to generate undetectable attacks. For instance: the Num\_GET feature is increased only by 2.19 on average, meaning that the attacker needs to add only 2 or 3 additional GET requests to the domain; the number of sub domains is increased by 0.21; the registration age of the domain is increased by 1.23 days, and the update age is increased by 17.84 days on average.
Lastly, we  include $L_2$ norm statistics of perturbations  added to create adversarial examples by the Penalty attack at distance 5 in the third row of Table~\ref{tab:l2_perturb_stats_made}. We notice that the average $L_2$ norm of perturbation (2.76) is much smaller than the standard deviation of the training samples $L_2$ norm (26.9). The maximum $L_2$ norm of perturbation is only a 0.005 fraction of the maximum possible $L_2$ norm of samples in the training data, confirming the fact that the resulting adversarial examples can be considered stealthy. Perturbations with the Penalty attacks are slightly lower than those generated by the Projected attack.

\begin{table}[ht]
\centering
\begin{tabular}{|c||c|}
\hline
Feature & Average perturbation\\ 
\hline

Total\_Recv\_Bytes & 76616.96\\
Total\_Sent\_Bytes & 8437.57\\
Min\_Ratio\_Bytes & 21.04\\
Num\_GET & 2.19\\
Sub\_Domains & 0.21\\
Reg\_Age & 1.23\\
Reg\_Validity & 113.51\\
Update\_Age & 17.84\\
Update\_Validity & 48.93\\

\hline
\end{tabular}
\caption{Average modified features perturbation for the \system\ Penalty attack at distance 20.}
\label{tab:pen_ave_features}
\end{table}

\begin{figure}[ht]
\begin{center}
\includegraphics[width=2.2in]{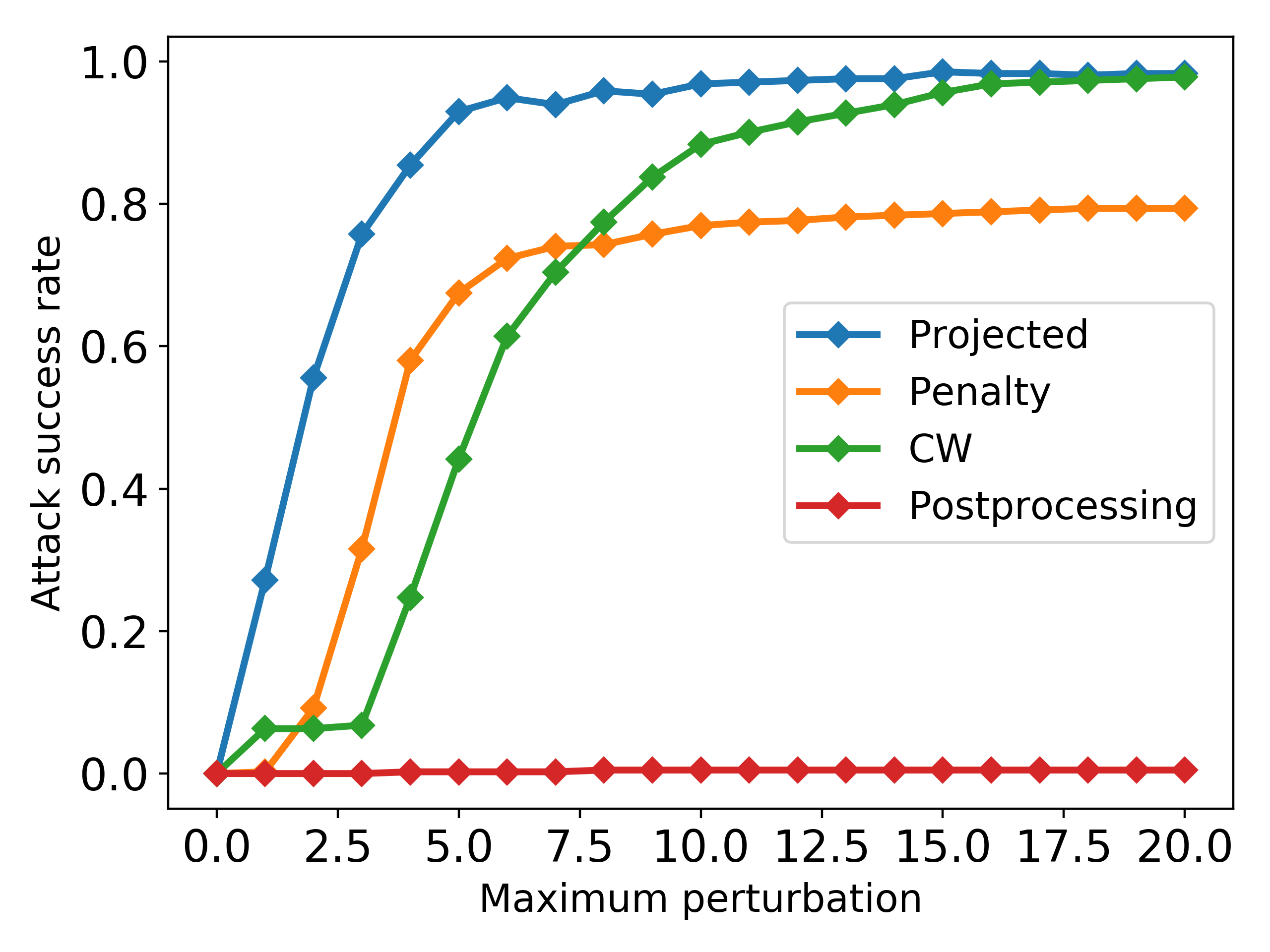}
\caption{Malicious domain classifier attacks. \system\ Projected attacks perform best. C\&W does not generate feasible adversarial examples.}
\label{fig:attacks_sr}
\end{center}
\end{figure}

\myparagraph{Attack comparison} We compare the success rate of our Projected and Penalty \system\ attacks with the C\&W attack, as well as an attack we call Post-processing. The Post-processing attack runs directly the original C\&W developed for continuous domains, after which it projects the adversarial example to the raw input space to enforce the constraints. For each family of dependent features, the attack retains the value of the representative feature but then modifies the  dependent features using the \updatedep\ function. The success rate of all these attacks is shown in Figure~\ref{fig:attacks_sr}, using the 412 \mal\ testing examples classified correctly. The attacks based on our \system\ framework (with Projected and Penalty objectives) perform best, as they account for feature dependencies during the adversarial example generation. The attack with the Projected objective has the highest performance.  The vanilla C\&W has slightly worse performance at small perturbation values, even though it does not take into consideration the feature constraints and works in an enlarged feature space. Interestingly, the Post-processing attack performs worse (reaching only 0.005\% success at a distance of 20 -- can generate 2 out of 412 adversarial examples). This demonstrates that it is not sufficient to run state-of-the-art attacks for continuous domains and then adjust the feature dependencies, but more sophisticated attack strategies are needed.

\myparagraph{Number and importance of features modified} 
We compare how many features were modified in order to generate each of the three attacks: Projected, Penalty, and C\&W.

It is not surprising that the C\&W attack  modifies almost all features, as it works in $L_2$  norms without any restriction in feature space. Both the Projected and the Penalty attacks modify a much smaller number of features (4 on average).

We are interested in determining if there is a relationship between feature importance and choice of feature by the optimization algorithm. For additional details on feature description, we include the list of features that can be modified in Table~\ref{tab:features_id}. 

We observe that features of higher importance are chosen more frequently by the optimization attack. However, since we are modifying the representative feature in each family, the number of modifications on the representative feature is usually higher (it accumulates all the importance of the features in that family). For the Bytes family, feature 3 (number of received bytes) is the representative feature and it is updated more than 350 times. However, for features that have no dependencies (e.g., 68 -- number of levels in the domain, 69 -- number of sub-domains, 71 -- domain registration age, and 72 -- domain registration validity), the number of updates corresponds to the feature importance.

\begin{figure*}[t]
\centering
  \begin{subfigure}[b]{0.37\linewidth}
    \includegraphics[width=\linewidth]{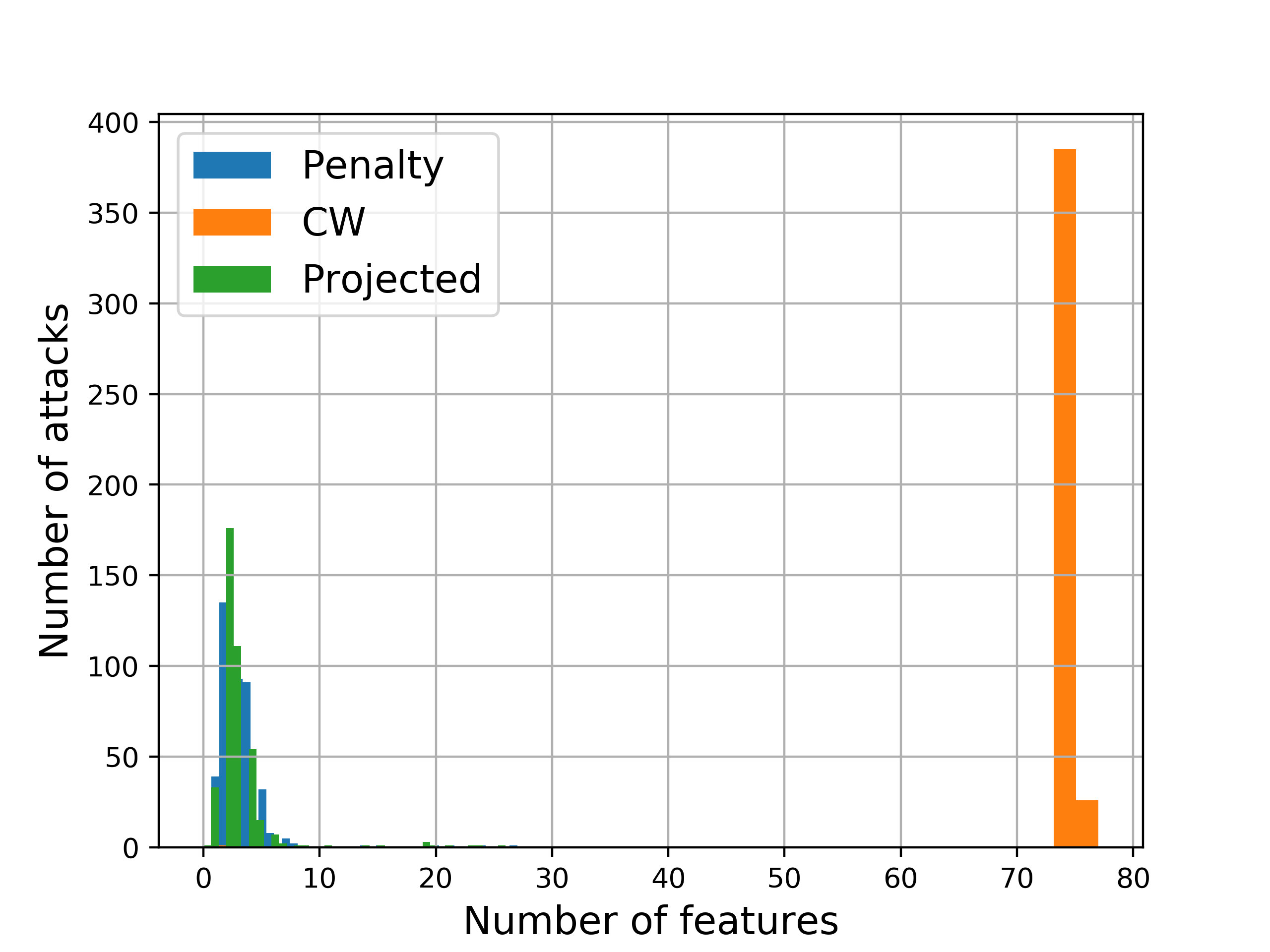}
    \caption{Histogram on feature\\ modifications.}
    \label{fig:features}
  \end{subfigure}
  \begin{subfigure}[b]{0.57\linewidth}
    \includegraphics[width=\linewidth]{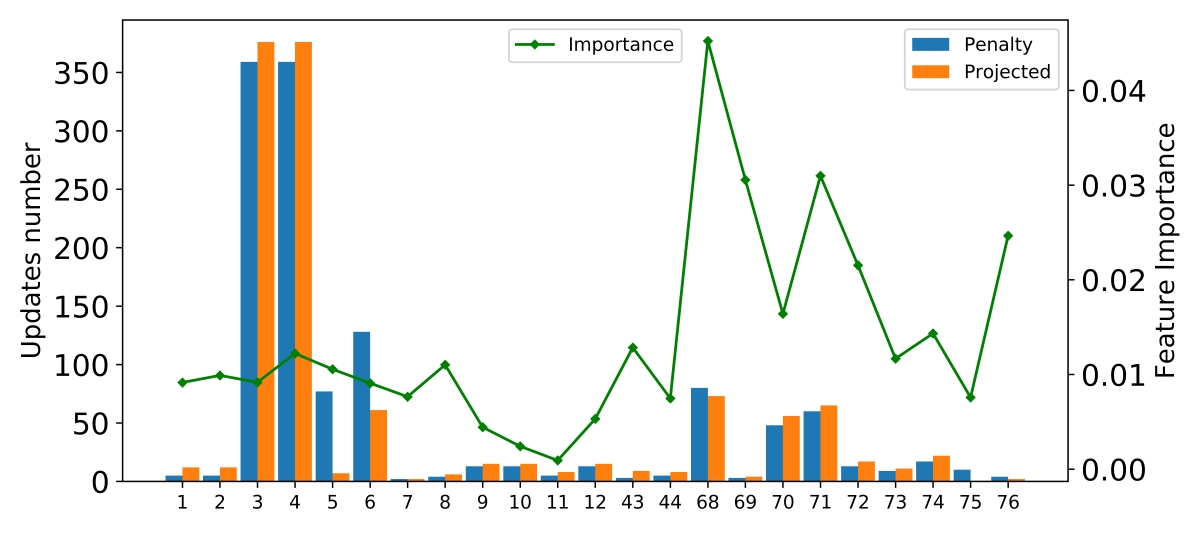}
    \caption{Number of updates (left),\\ feature importance (right).}
     \end{subfigure}
\caption{Feature modification statistics for malicious domain classifier.}
\label{fig:f_times}
\end{figure*}

\subsection{Misclassification of \ben\ Samples as \mal} 

\begin{figure*}[ht]
\centering
  \begin{subfigure}[b]{0.4\linewidth}
    \includegraphics[width=\linewidth]{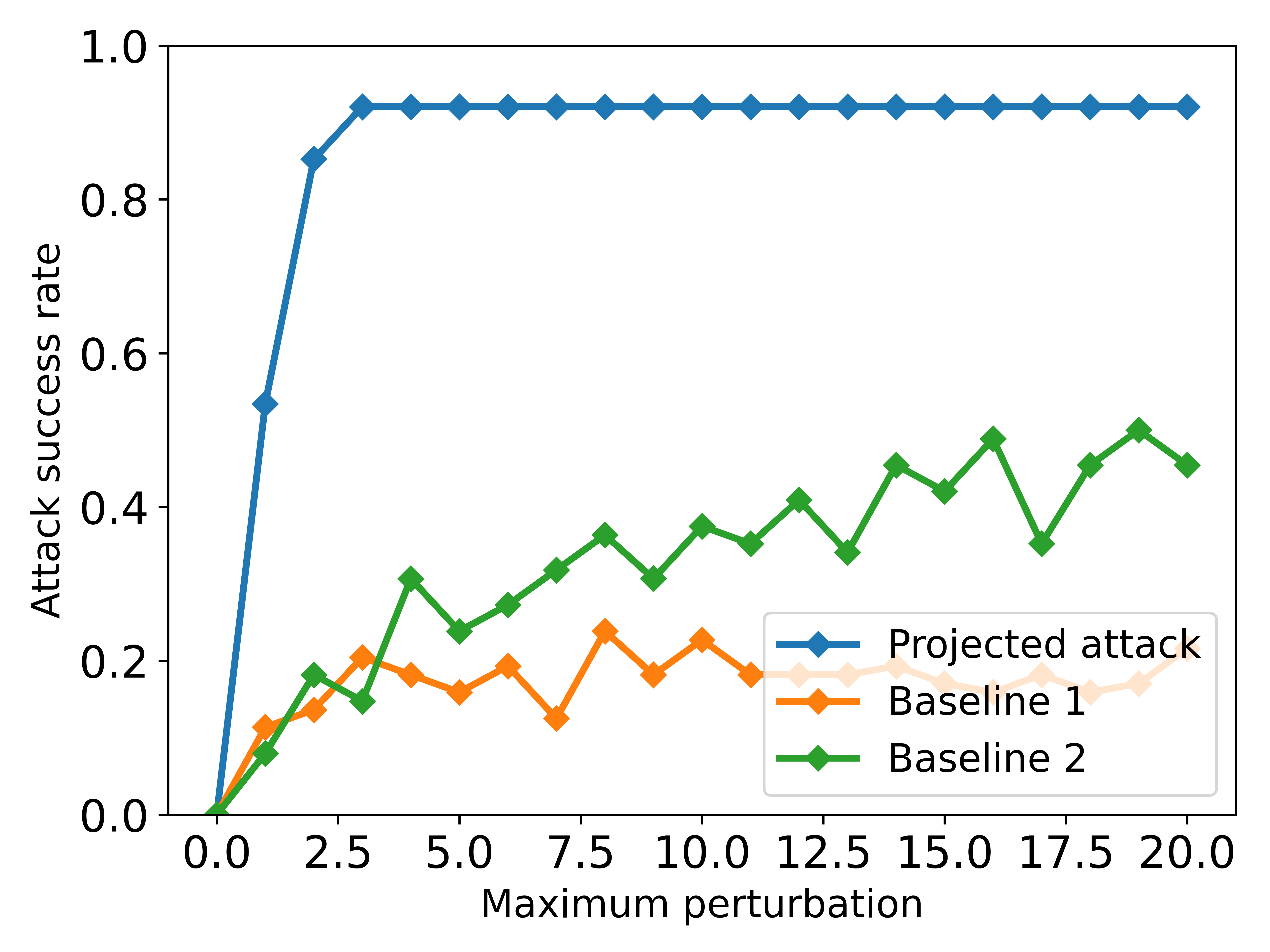}
    \caption{Projected attack.}
    \label{fig:models}
  \end{subfigure}
  \begin{subfigure}[b]{0.4\linewidth}
    \includegraphics[width=\linewidth]{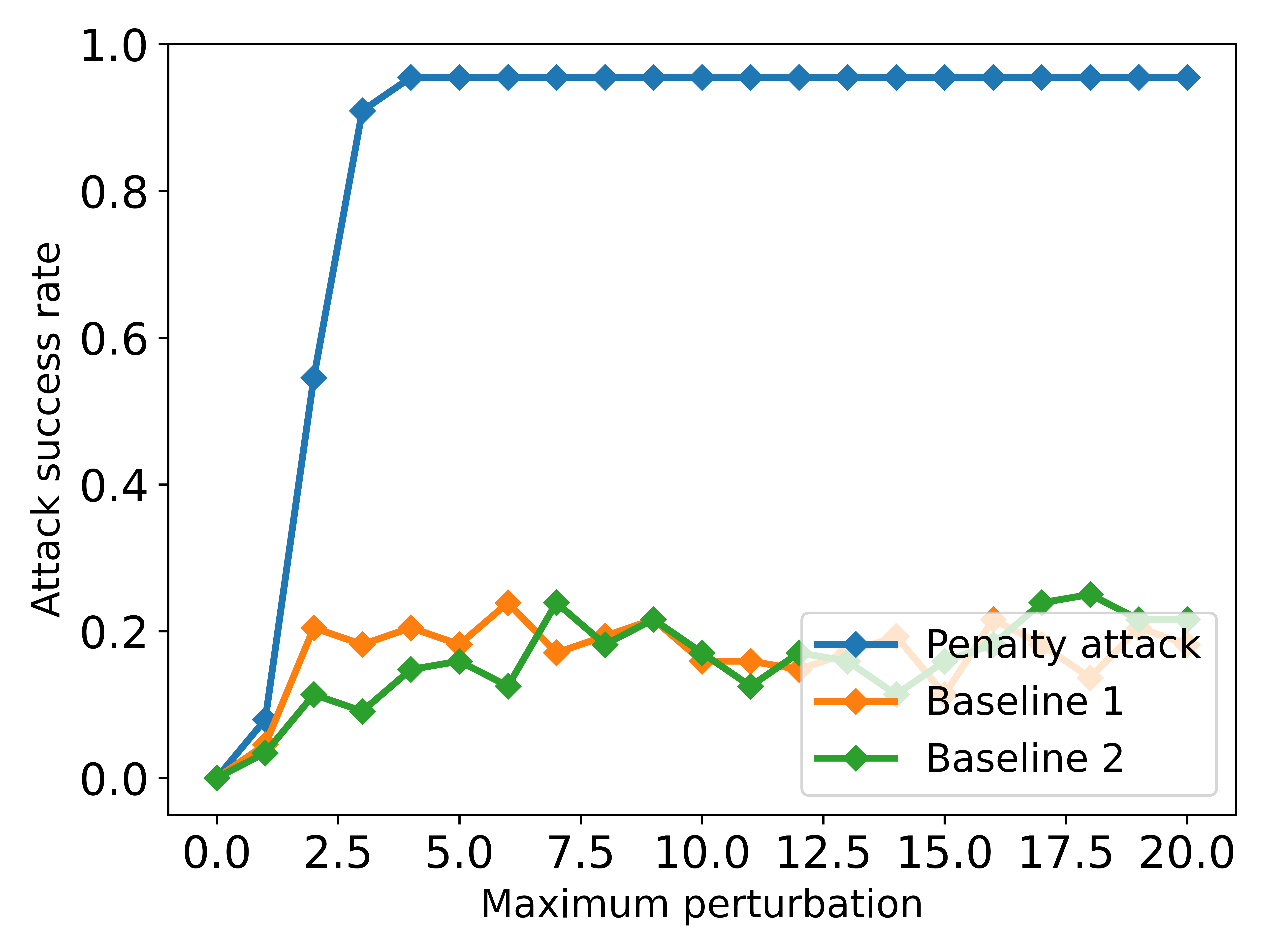}
    \caption{Penalty attack.}
  \end{subfigure}
\caption{\system\ attack results for misclassification of \ben\ samples to \mal.}
\label{fig:ben_to_mal}
\end{figure*}

We ran the Projected and Penalty attacks for changing the classification of \ben\ inputs to \mal\ to show the generality of the \system\ framework. The success rate of the attacks compared to two  baseline attacks for balanced classes is illustrated in Figure~\ref{fig:ben_to_mal}. Both attacks achieve high success rate: At $L_2 = 20$ the Projected attack reaches 92\% success rate and the Penalty attack reaches 95\% success rate. The \system\ attack performs significantly better compared to both baseline attacks.

\subsection{Attack Transferability}  We consider here a threat model in which the adversary only knows the feature representation, but not the exact ML model or the training data. One approach to generate adversarial examples is through transferability~\cite{papernot2016transfer,song2016transfer, tramer2017transfer,suciu2018transfer,demontis2019transfer}. We perform several experiments to test the transferability of the Projected attacks against \ffnn\ to logistic regression (LR) and random forest (RF). Models were trained with different  data and we vary the imbalance ratio. The results are in Table~\ref{tab:transferfromdnn}. We observe that the largest transferability rate to both LR and RF is for the highest imbalanced ratio of 50 (98.2\% adversarial examples transfer to LR and 94.8\% to RF). As we increase the imbalance ratio, the transfer rate increases and the transferability rate to LR is lower than to RF. 

\begin{table}[ht]
\centering
\begin{tabular}{|c||c||c||c|}
\hline
Ratio& \ffnn\ & LR & RF\\ 
\hline
1&100\%&40\%&51.7\%\\
5&93.3\%&66.5\%&82.9\%\\
15&99\%&60.9\%&90.2\%\\
25&100\%&47.6\%&68.8\%\\
50&100\%&98.2\%&94.8\%\\
\hline
\end{tabular}
\caption{Transferability of adversarial examples from \ffnn\ to LR  and RF. We vary the imbalance ratio in training. Column \ffnn\ shows the white-box attack success rate.}
\label{tab:transferfromdnn}
\end{table}

We also look at the transferability between different \ffnn\ architectures trained on different datasets (results in Table~\ref{tab:transferbetweendnn}). The attacks transfer best at the highest imbalance ratio (with a success rate higher than 96\%), confirming that weaker models are easier to attack.

\begin{table}[ht]
\centering
\begin{tabular}{|c||c||c||c|}
\hline
Ratio& DNN1 & DNN2 & DNN3\\ 
& [80, 50] & [160, 80] & [100, 50, 25] \\
\hline
1&100\%&57.6\%&42.3\%\\
5&93.3\%&73.6\%&58.6\%\\
15&99\%&78.6\%&52.4\%\\
25&100\%&51.4\%&45.3\%\\
50&100\%&96\%&97.1\%\\
\hline
\end{tabular}
\caption{Transferability between architectures (number of neurons per layer in the second row). Adversarial examples  computed for DNN1 are transferred to DNN2 and DNN3.}
\label{tab:transferbetweendnn}
\end{table}

\subsection{Mitigations} \label{sec:advtr}
\begin{figure*}[t]
\begin{center}
\includegraphics[width=2.2in]{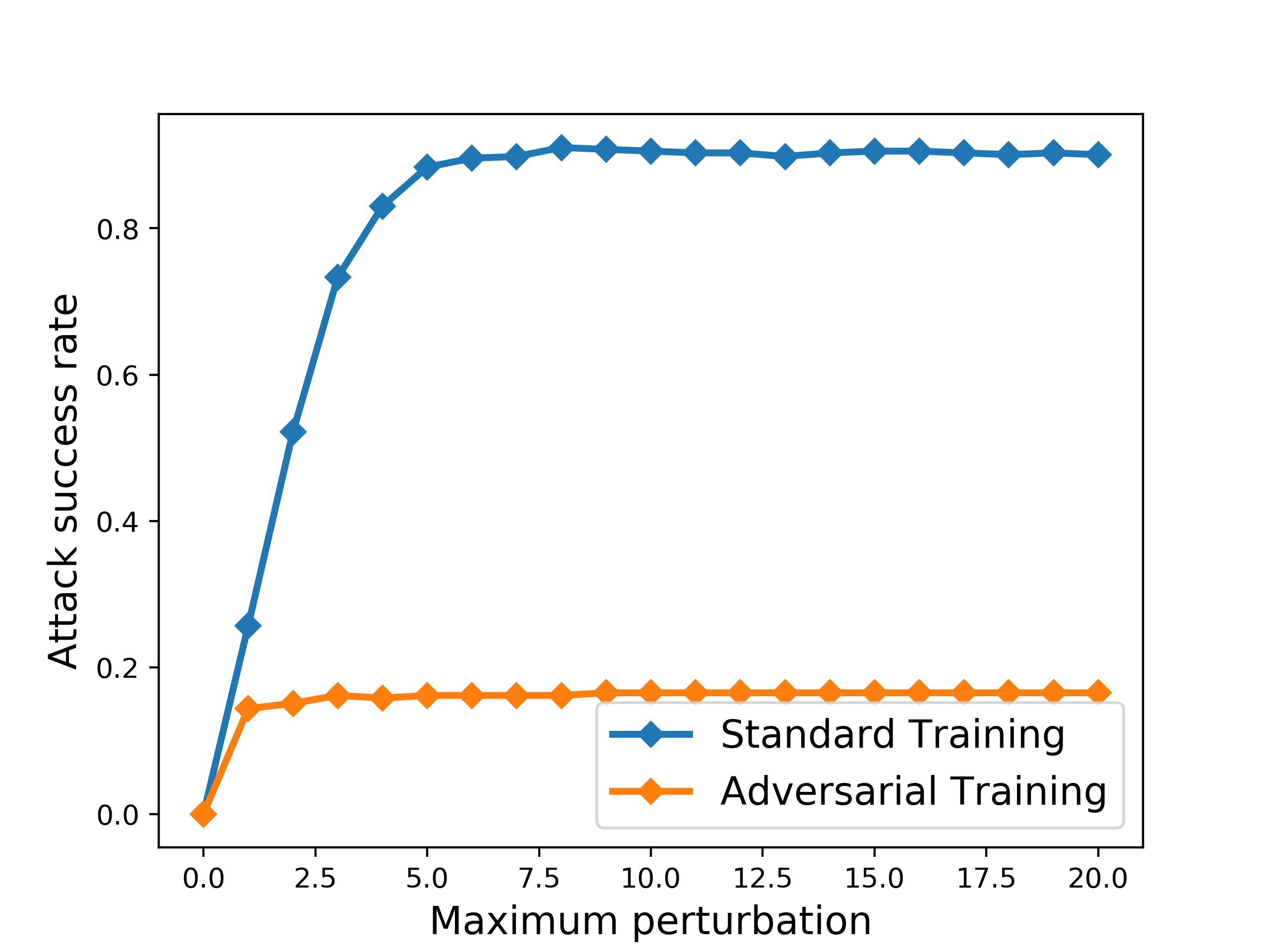}
\caption{Success rate of the FENCE Projected attack against adversarially and standard trained model.}
\label{fig:adv_training}
\end{center}
\end{figure*}

Finally, we looked at defensive approaches to increase the \ffnn\ robustness against the \system\ evasion attack. A well-known defensive technique  is adversarial training~\cite{Goodfellow14,madry2017towards}. We trained \ffnn\ using adversarial training with the Projected attack at $L_2$ distance 20.  We trained the model adversarially for 11 epochs and obtained the AUC score of 89\% (each epoch takes approximately 7 hours).  We measured the Projected attack's success rate for the balanced case against the standard and adversarially training models in Figure~\ref{fig:adv_training}. Interestingly, the success rate of the evasion attacks significantly drops for the adversarially-trained model and reaches only 16.5\% at 20 $L_2$ distance. This demonstrates that adversarial training is a promising direction for designing robust ML models for security. We plan to investigate it further and optimize its design in future work.

\section{Related Work}
\label{sec:related}

Adversarial machine learning  studies ML vulnerabilities against attacks~\cite{Huang2011adversarial}. Research on the robustness of DNNs at testing time started with the work of Biggio et al.~\cite{Biggio13} and Szegedy et al.~\cite{Szegedy14}. They showed that classifiers are vulnerable to adversarial examples generated with minimal perturbation to testing inputs. Since then, the area of adversarial ML has received a lot of attention, with the majority of work focused on evasion attacks (at testing time), e.g.,~\cite{Goodfellow14,kurakin2016adversarial,Papernot17,Papernot-BlackBox17,Carlini17,athalye2018obfuscated,lujoreiter}. Other classes of attacks include poisoning (e.g.,~\cite{biggio2012poisoning,Xiao15}) and privacy attacks (e.g.,~\cite{Fredrikson15,Membership}), but we focus here on evasion attacks.

\myparagraph{Evasion attacks in security}
Several evasion attacks have been proposed against models with discrete and constrained input vectors, as encountered in security. The majority of these use datasets with binary features, not considering dependencies in feature space. Biggio  et  al.~\cite{Biggio13} use  a  gradient-based  attack to construct adversarial examples for malicious PDF detection by only adding new keywords to PDFs. Grosse et al.~\cite{grosse2016adversarial} leverage the JSMA attack by Papernot et al.~\cite{Papernot17} for a malware classification application in which features can be added or removed. Suciu et al.~\cite{suciu2018exploring} add bytes to malicious binaries either at the end or in slack regions to create adversarial examples. Kreuk~\cite{kreuk2018deceiving} discover regions in executables  that would  not  affect  the  intended  malware behavior. Kolosnjaji  et  al.~\cite{kolosnjaji2018adversarial} create gradient-based attack against malware  detection   DNNs that learn from  raw  bytes, and can create adversarial examples by only changing a few  specific  bytes  at  the  end  of  each  malware  sample. Xu et al.~\cite{xu2016automatically} propose a black-box attack based on  genetic algorithms for manipulating  PDF files while  maintaining  the required format. Dang et al.~\cite{dang2017evading} propose a black-box attack against  PDF malware classifiers that uses hill-climbing over a set of feasible transformations. Anderson et al.~\cite{anderson2018learning} construct a general black-box framework based on reinforcement learning for attacking static portable executable anti-malware engines. Kulynych et al.~\cite{kulynych2018evading} propose a graphical  framework  for discrete domains with guarantees of minimal adversarial cost. Recently, Pierazzi et al.~\cite{pierazzi2019intriguing} define a formalization for the domain-space attacks, along with a new white-box attack against Android malware classification. The authors use automated software transplantation to extract slices of bytecode from benign applications and inject them into a malicious host to mimic the benign activity and evade the classifier.
Chen et al. ~\cite{chen2017adversarial} proposed an evasion attack called EvnAttack on malware present in portable Windows executable files, where input vectors are binary features each representing an API call to Windows.

Evasion attacks for network traffic classifiers include: Apruzesse et al.~\cite{apruzzese2018} analyzing the robustness of random forest for botnet classification; Clements et al.~\cite{clements2019rallying}  evaluating the robustness of an anomaly detection method~\cite{mirsky2018kitsune} against existing attacks; and De Lucia et al.~\cite{de2019adversarial} attacking an SVM for network scanning detection. A number of papers perform attacks against intrusion detection systems (IDS). Among them, Warzynski and Kołaczek
~\cite{warzynski2018intrusion} consider an FGSM attack under L1 norm against IDS and illustrates the ability of the generated adversarial example to evade the classifier. Rigaki et al. ~\cite{rigaki2017adversarial} generate targeted attacks by using FGSM and JSMA to evade decision tree, random forest, linear SVM, voting ensembles of the previous three classifiers, and a multi-layer perceptron (MLP) neural network IDS. Similarly, Wang et al ~\cite{wang2018deep} leveraged FGSM, JSMA, Deepfool, and Carlini-Wagner to attack an MLP neural network for intrusion detection. Yang et al. ~\cite{yang2018adversarial} used Carlini-Wagner, a GAN attack, and black-box ZOO attacks against IDS DNNs. In their work Martins et al.~\cite{martins2019analyzing} tested the performance of FGSM, JSMA, Deepfool, and Carlini-Wagner attacks against decision tree, random forest, SVM, naive Bayes, neural networks, and denoising autoencoders intrusion detection models. Wu et al. ~\cite{wu2019evading} applied deep reinforcement learning to generate adversarial attacks on botnet attacks. Yan et al. ~\cite{yan2019automatically} generate adversarial examples for denial of service attacks. Lin et al. ~\cite{lin2018idsgan} use a modification of GAN called IDSGAN to produce adversarial examples while retaining functional features of the attack. \\

\noindent {\bf Evasion attacks with dependency constraints.} A number of attacks that preserve constraints between extracted features in security exist in the literature. We survey these papers in greater detail and add a comparison to \system\ in Table~\ref{tab:refs}.

Alhajjar et al.~\cite{alhajjar2020adversarial}  explore the use of evolutionary computations and generative adversarial networks as a tool for crafting adversarial examples that aim to evade machine learning models used for network traffic classification. These strategies were applied to the NSL-KDD and UNSW-NB15 datasets. The paper operates only in  features space, and the following dependencies are preserved: binary features, linear dependencies  and features that can only be increased. There is no limitation on the amount of perturbations, as the authors claim that large changes in data are not subject to easy recognition by human observers. 
The percent of successful evasion for multi-layer perceptron (MLP) is much  worse for both datasets than in \system: maximum 86.18\% for NSL-KDD using GANs, and 55.10\% for the UNSW-NB15 dataset.
In contrast, \system\ minimizes the perturbation needed for evasion, preserves a much larger number of dependencies, such as non-linear, statistical, or combination of them, achieving a higher success rate. 

Granados et al.~\cite{granados2020realistic}  introduce Restricted Traffic Distribution Attack (RTDA) against network traffic classifiers, which is based on the Carlini and Wagner attack. In order to ensure the feasibility of adversarial examples, the attack only modifies the features corresponding to different percentiles for packet sizes by increasing their values and preserving the monotonic non-decreasing property of generated packet-size distribution. 
The attacker operates only in  feature space, and there is no limit on bytes added to the packet, which may result in the packet sizes that are larger than allowed by network protocols. In contrast, \system\  generates  more realistic attacks in the real-world scenario, by adding new connections of different types (UDP/TCP) and on different ports. Moreover, \system\ ensures feasibility by preserving packet sizes, connection durations, and other network traffic characteristics according to the constraints of network protocols.

Abusnaina et al.~\cite{abusnaina2019examining}  present MergeFlow attack against DDoS detection models. They create adversarial examples by combining features from existing network flow with a representative mask flow from the target class. The features are combined either through averaging for ratio features or accumulating for count-based features. This attack preserves dependencies between features but generates large perturbations.
Abusnaina et al. present similar ideas against graph-based IoT Malware Detection System that creates realistic adversarial examples by combining the original graph with a selected targeted class.

Sadeghzadeh et al.~\cite{sadeghzadeh2021adversarial}  introduce an adversarial network traffic attack (ANT) that uses a universal adversarial perturbation (UAP) generating method. To generate ANT they introduce the following three types of attack: AdvPad injects UAP to the content of packets to evaluate the robustness of packet classifiers, AdvPay injects UAP into payload if a dummy packet to evaluate flow-based classifiers, and AdvBurst  modifies a burst of the flow to test the robustness of flow time-series classifiers. In order to generate the UAP, a set of flows or packets from a particular class is leveraged and the UAP is inserted into new incoming traffic of that class. This attack works in the domain space by inserting payload, dummy packet with payload, or the sequence of packets with statistical features. All perturbations are calculated by inserting the randomly initialized perturbation vector into examples from the target class and optimizing the loss function towards the selected class for some number of iterations. Sequences of bytes were used for training the classifiers, thus, there are no dependencies in the domain space. The only thing that needs to be preserved is the range for the number of bytes achieved by performing the clipping operation while computing UAP.

Han et al.~\cite{han2020practical} propose a practical traffic-space evasion attack against Network Intrusion Detection System (NIDS). The attacker can modify the original traffic generated from devices he controls at an affordable overhead. It has two main steps: finding the adversarial feature vector which can be classified as benign but is close to the malicious feature vector in terms of features' values (lies in the low-confidence region of the classifier)  using the GAN model and transforming malicious original traffic to transfer its features to the closest adversarial ones preserving its functionality using particle swarm optimization (PSO). While performing the second step authors allow only to modify the interarrival time of packets in the original traffic, and for the injected crafted traffic the attacker is able to alter the interarrival time of packets, protocol layer of packets, and payload size. The budget overhead for the number of buckets and time elapsed is controlled by the rate of the original traffic. Feature dependencies are preserved in the domain space only, in contrast, \system\ allows to additionally preserve complex mathematical dependencies in  feature space.

Chen et al. ~\cite{chen2020generating} introduce two types of evasion attacks against machine learning-based intrusion detection systems. The first attack called $Opt$ uses an iterative optimization approach to find the adversarial examples that maximize the probability of malicious output while remaining stealthy and preserving the Range dependency. The second attack detaches the malicious payload from the input vector, and then optimizes the modifiable features based on a GAN under the Range dependency, keeping some features as integers. Thus, both attacks preserve only domain-specific dependencies, while \system\ is capable of satisfying a larger amount of domain dependencies as well as a list of mathematical constraints.

None of these previous works can handle the same amount of  domain and mathematical  dependencies supported by our \system\ framework.
In Table~\ref{tab:refs} we compare \system\ with these attacks against network traffic classifiers and include the list of supported constraints.
 
\begin{table}[ht]
\centering
\begin{tabular}{|c||c||c||c||c|}
\hline
Paper & Classifiers & Algorithm & Domain & Mathematical \\ 
& & & Dependencies & Dependencies \\
\hline
FENCE&Network Traffic, &Iterative &Ratio&Stat\\
&Malicious Domains&optimization&Range&Lin\\
& & Projected & OHE & Non-Lin\\
& & Penalty &&Combination\\
\hline
Alhajjar et al.~\cite{alhajjar2020adversarial}&Network Traffic&GAN/PSO&OHE&Linear\\
\hline
Granados et al. ~\cite{granados2020realistic}&Network Traffic&Iterative&Range&Stat\\
&&optimization&&\\
\hline
Sadeghzadeh et al.~\cite{sadeghzadeh2021adversarial}&Packets&UAP&Range&-\\
&Network Flows&&&\\
&Network Bursts&&&\\
\hline
Han et al. ~\cite{han2020practical}&Botnet&GAN/PSO&Range&-\\
\hline
Abusnaina et al.~\cite{abusnaina2019examining}&DDoS Detection&Sample&Range&-\\
&&injection&Ratio&-\\
\hline
Chen et al.~\cite{chen2020generating}&Network Traffic&Iterative&Range&-\\
&&optimization&&-\\
&&GAN&&\\
\hline
\end{tabular}
\caption{Comparison to existing work on evasion attacks in cybersecurity domains. We only include methods which respect dependencies in feature space. For each method, we mention the adversarial attack algorithm and the supported feature dependencies.}
\label{tab:refs}
\end{table}

\myparagraph{Evasion attacks in other domains}
There is work on designing attacks in other domains, such as audio:
~\cite{gong2017crafting},
~\cite{cisse2017houdini},
~\cite{zhang2017dolphinattack},
~\cite{song2017inaudible},
~\cite{schonherr2018adversarial},
~\cite{yakura2018robust},
~\cite{carlini2018}
~\cite{qin2019imperceptible};
 text:
~\cite{papernot2016crafting}, 
~\cite{ebrahimi2017hotflip}, 
~\cite{liang2017deep}, 
~\cite{gao2018black}, 
~\cite{alzantot2018generating}; 
and video:
~\cite{li2019stealthy},
~\cite{hosseini2017attacking},
~\cite{wei2018sparse}.  Physically realizable attacks have been designed for face recognition~\cite{Sharif16} and vision~\cite{RobustPA}.

\vspace{0.1cm}

\myparagraph{Defenses against evasion attacks} Standard methods to defend against adversarial evasion attacks in continuous domains include: adversarial training~\cite{madry2017towards},  randomized smoothing~\cite{cohen2019certified}, and defenses based on the detection of adversarial examples~\cite{roth2019odds, li2019generative, yin2019adversarial, yu2019new, yang2018characterizing}. Randomized smoothing provides certifiable guarantees, but unfortunately, it cannot be directly applied in discrete domains because it requires the classifier to be evaluated on inputs perturbed with Gaussian noise. Therefore, we are not aware of any uses of randomized smoothing and similar randomization-based mechanism for cyber security defenses. 

Defenses based on adversarial training have been used in cyber security. For instance, Abusnaina et al.~\cite{abusnaina2019examining} validate whether adversarial training improves the robustness of the DDoS detection system. They show that adversarial training is able to improve the robustness of the anomaly classification model, but only for certain types of adversarial examples included in the adversarial training. Hashem et al.~\cite{hashemi2020enhancing}  develop a network intrusion detection system that utilizes a novel reconstruction from partial observation method to build a more accurate anomaly detection model. The method also improves robustness to adversarial examples.

Methods for the detection of adversarial examples in security have been proposed. For example, Wang et al. ~\cite{wang2021manda}  propose a defense mechanism against adversarial examples for NIDS, which exploits the inconsistency between manifold evaluation and model inference along with the closeness of the resulting adversarial example to the manifold. This allows them to train a logistic regression that predicts whether the input may be considered adversarial or not. In general, a known limitation of detection methods is that they fail in face of adaptive attacks. Tramer et al.~\cite{tramer2020adaptive} showed that multiple detection-based methods (e.g.~\cite{roth2019odds}) are not resilient against adaptive attacks, and thus designing detection-based defenses in cyber security which are resilient to adaptive attacks remains an open problem.

\section{Conclusions}
\label{sec:conclusion}

We  showed  that evasion  attacks  against DNNs can be designed to preserve the dependencies in feature space  in constrained domains. We proposed a general framework \system\ for  generating adversarial examples that respects mathematical dependencies and domain-specific constraints imposed by these applications. We demonstrated evasion attacks that insert a small number of network connections (12 records in Zeek connection logs) to misclassify \mal\ activity as \ben\ in a malicious connection classifier. We also showed that adversarial training has the potential to increase the robustness of classifiers in the malicious domain setting. 

\noindent Our \system\ framework is not restricted to  security applications, and we plan to apply it to  healthcare and financial scenarios. An important open problem in this space is how to increase the resilience of DNN  classifiers used in critical, constrained applications.
\begin{acks}
We thank Simona Boboila and Talha Ongun for generating the features used for the malicious network traffic classifier. 
This project was funded by NSF under grant CNS-1717634 and by a Google Security and Privacy Award. This research was also sponsored by the U.S. Army Combat Capabilities Development
Command Army Research Laboratory under Cooperative Agreement Number W911NF-13-2-0045 (ARL Cyber Security CRA), and by the contract number W911NF-18-C0019 with the U.S. Army Contracting Command - Aberdeen
Proving Ground (ACC-APG) and the Defense Advanced Research Projects Agency (DARPA). The views and conclusions contained in this document are those of the authors and should not be interpreted as representing the official policies, either expressed or implied, of the Combat Capabilities Development Command Army Research Laboratory, ACC-APG, DARPA, or the U.S. Government. The U.S. Government is authorized to reproduce and distribute reprints for Government purposes notwithstanding any copyright notation here on.
\end{acks}

\bibliographystyle{ACM-Reference-Format}
\bibliography{refs}

\end{document}